\documentclass[twocolumn]{aastex631}

\usepackage{multirow}
\usepackage{amsmath}
\usepackage{hyperref}

\newcommand{\sci}[2]{#1 $\times$ 10$^{#2}$}

\received{April 10, 2021}
\accepted{June 27, 2021}
\submitjournal{ApJS}

\shorttitle{Massive compact disks around FUors revealed by ALMA}
\shortauthors{K\'osp\'al et al.}

\begin{document}

\title{Massive compact disks around FU Orionis-type young eruptive stars revealed by ALMA}

\author[0000-0001-7157-6275]{\'A. K\'osp\'al}
\affiliation{Konkoly Observatory, Research Centre for Astronomy and Earth Sciences, E\"otv\"os Lor\'and Research Network (ELKH), Konkoly-Thege Mikl\'os \'ut 15-17, 1121 Budapest, Hungary}
\affiliation{Max Planck Institute for Astronomy, K\"onigstuhl 17, 69117 Heidelberg, Germany}
\affiliation{ELTE E\"otv\"os Lor\'and University, Institute of Physics, P\'azm\'any P\'eter s\'et\'any 1/A, 1117 Budapest, Hungary}
  \email{kospal@konkoly.hu}

\author[0000-0002-4283-2185]{F. Cruz-S\'aenz de Miera}
  \affiliation{Konkoly Observatory, Research Centre for Astronomy and Earth Sciences, E\"otv\"os Lor\'and Research Network (ELKH), Konkoly-Thege Mikl\'os \'ut 15-17, 1121 Budapest, Hungary}

\author[0000-0001-8445-0444]{J. A. White}
  \affiliation{Jansky Fellow of the National Radio Astronomy Observatory.; National Radio Astronomy Observatory, 520 Edgemont Road, Charlottesville, VA 22903, USA}
  \affiliation{Konkoly Observatory, Research Centre for Astronomy and Earth Sciences, E\"otv\"os Lor\'and Research Network (ELKH), Konkoly-Thege Mikl\'os \'ut 15-17, 1121 Budapest, Hungary}

\author[0000-0001-6015-646X]{P. \'Abrah\'am}
\affiliation{Konkoly Observatory, Research Centre for Astronomy and Earth Sciences, E\"otv\"os Lor\'and Research Network (ELKH), Konkoly-Thege Mikl\'os \'ut 15-17, 1121 Budapest, Hungary}
\affiliation{ELTE E\"otv\"os Lor\'and University, Institute of Physics, P\'azm\'any P\'eter s\'et\'any 1/A, 1117 Budapest, Hungary}

\author{L. Chen}
  \affiliation{Konkoly Observatory, Research Centre for Astronomy and Earth Sciences, E\"otv\"os Lor\'and Research Network (ELKH), Konkoly-Thege Mikl\'os \'ut 15-17, 1121 Budapest, Hungary}

\author[0000-0002-6018-1371]{T. Csengeri}
  \affiliation{Max-Planck-Institut f\"ur Radioastronomie, Auf dem H\"ugel 69, 53121 Bonn, Germany}

\author[0000-0001-9290-7846]{R. Dong}
  \affiliation{Department of Physics \& Astronomy, University of Victoria, Victoria, BC, V8P 1A1, Canada}

\author[0000-0003-0749-9505]{M. M. Dunham}
  \affiliation{Department of Physics, State University of New York at Fredonia, Fredonia, NY 14063, USA}

\author[0000-0003-3453-4775]{O. Feh\'er}
  \affiliation{Konkoly Observatory, Research Centre for Astronomy and Earth Sciences, E\"otv\"os Lor\'and Research Network (ELKH), Konkoly-Thege Mikl\'os \'ut 15-17, 1121 Budapest, Hungary}
  \affiliation{IRAM, 300 Rue de la piscine, 38406 Saint-Martin-d’H\`eres, France}

\author[0000-0003-1665-5709]{J. D. Green}
  \affiliation{The University of Texas at Austin, Department of Astronomy, 2515 Speedway, Stop C1400, Austin, TX 78712, USA}
  \affiliation{Space Telescope Science Institute, 3700 San Martin Dr., Baltimore, MD 02138, USA}

\author[0000-0002-3053-3575]{J. Hashimoto}
  \affiliation{Astrobiology Center of NINS, 2-21-1, Osawa, Mitaka, Tokyo, 181-8588, Japan}
  
\author[0000-0002-1493-300X]{Th. Henning}
  \affiliation{Max Planck Institute for Astronomy, K\"onigstuhl 17, 69117 Heidelberg, Germany}

\author[0000-0001-5217-537X]{M. Hogerheijde}
  \affiliation{Leiden Observatory, Leiden University, P.O. Box 9513, 2300 RA Leiden, The Netherlands}
  \affiliation{Anton Pannekoek Institute for Astronomy, University of Amsterdam, Science Park 904, 1098 XH Amsterdam, The Netherlands}

\author[0000-0002-9294-1793]{T. Kudo}
  \affiliation{Subaru Telescope, National Astronomical Observatory of Japan, National Institutes of Natural Sciences (NINS), 650 North A‘oh\=ok\=u Place, Hilo, HI 96720, USA}

\author[0000-0003-2300-2626]{H. B. Liu}
  \affiliation{11F of AS/NTU Astronomy-Mathematics Building, No.1, Sec. 4, Roosevelt Rd, Taipei 10617, Taiwan, R.O.C.}

\author[0000-0001-9248-7546]{M. Takami}
  \affiliation{11F of AS/NTU Astronomy-Mathematics Building, No.1, Sec. 4, Roosevelt Rd, Taipei 10617, Taiwan, R.O.C.}

\author[0000-0002-6045-0359]{E. I. Vorobyov}
  \affiliation{Department of Astrophysics, The University of Vienna, Vienna, A-1180, Austria}
  \affiliation{Research Institute of Physics, Southern Federal University, Stachki 194, Rostov-on-Don, 344090, Russia}


\begin{abstract}
FU Orionis-type objects (FUors) are low-mass pre-main sequence stars undergoing a temporary, but significant increase of mass accretion rate from the circumstellar disk onto the protostar. It is not yet clear what triggers the accretion bursts and whether the disks of FUors are in any way different from disks of non-bursting young stellar objects. Motivated by this, we conducted a 1.3\,mm continuum survey of ten FUors and FUor-like objects with ALMA, using both the 7\,m array and the 12\,m array in two different configurations to recover emission at the widest possible range of spatial scales. We detected all targeted sources and several nearby objects as well. To constrain the disk structure, we fit the data with models of increasing complexity from 2D Gaussian to radiative transfer, enabling comparison with other samples modeled in a similar way. The radiative transfer modeling gives disk masses that are significantly larger than what is obtained from the measured millimeter fluxes assuming optically thin emission, suggesting that the FUor disks are optically thick at this wavelength. In comparison with samples of regular Class\,II and Class\,I objects, the disks of FUors are typically a factor of 2.9--4.4 more massive and a factor of 1.5--4.7 smaller in size. A significant fraction of them (65--70\%) may be gravitationally unstable.
\end{abstract}



\section{Introduction}

FU~Orionis-type objects (FUors) are low-mass pre-main sequence stars undergoing large outbursts in visible light, attributed to temporarily enhanced accretion \citep{hk96}. Episodes of high accretion are believed to occur at all early stages of star formation after the prestellar core phase, but become observable at optical or infrared wavelengths only as the circumstellar envelope thins \citep[][and references therein]{audard2014}. During outburst, accretion rate from the circumstellar disk to the star may reach up to $10^{-4}\,M_{\odot}$\,yr$^{-1}$, several orders of magnitude higher than in quiescence or in normal T\,Tauri stars. Repetitive FUor outbursts, each lasting $\sim$100 years and dumping as much as $10^{-2}\,M_{\odot}$ onto the star, significantly contribute to the build-up of the final stellar mass, and offer a possible solution for the long-standing ``protostellar luminosity problem'' \citep[e.g.,][]{dunham2014ppvi}. The physical mechanism of FUor outbursts is not yet known, but several processes have been proposed \citep{audard2014}.

FUor outbursts may eventually clear up the protostar's environment at the end of the embedded (Class\,I) phase. By this time, envelopes are depleted to a level comparable to the mass accreted in a single outburst, and UV/X-ray radiation and outflows accompanying the eruptions may disperse them altogether. Thus, individual FUor eruptions might modify the structure of the circumstellar environment considerably, both the envelope and the disk itself \citep{vorobyov2020}. Following a sequence of outbursts (``the FUor phase''), the envelope disappears, and the young star enters the disk-only Class\,II phase. FUor eruptions may also change the mineralogy and chemistry of the disk \citep{molyarova2018}. Therefore, FUor outbursts mark a short but important transitional period in the evolution of the circumstellar material \citep{takami2018, takami2019}. Numerical simulations support the notion that most FUors occur in the Class\,I or early Class\,II phases \citep{vorobyov2015}. Exploring this period via characterizing FUors, in particular their disks where the outburst happens, is the main topic of the present paper.

Millimeter continuum emission has long been used to study the dusty environment of FUors \citep{henning1998,sw2001}. Interferometric observations of the millimeter dust continuum are the standard way to estimate basic disk properties like their mass and size or characteristic radius. Using the Atacama Large Millimeter/submillimeter Array (ALMA), \citet{cieza2018} observed three FUors, four EXors (objects showing smaller, shorter, repetitive accretion outbursts), and one intermediate-type object. They found that FUor disks are significantly more massive than EXor disks, and they are more compact for a given disk mass than what is typical for normal T Tauri disks. \citet{liu2018} reached a similar conclusion in a Submillimeter Array (SMA) study of 29 FUors, EXors, and FUor-like objects. Overall, the observed targets showed a systematically higher millimeter luminosity distribution than those of normal T Tauri stars in nearby star-forming regions, and within the sample, all EXors had smaller 1.33\,mm fluxes (when scaled to the same distance) than any of the FUors. This suggests that relatively massive circumstellar disks may provide optimal conditions for accretion outbursts.  

Not all FUors are isolated objects, some of them are parts of small clusters of young stars (like AR~6A/6B, HBC~722, or V960~Mon, \citealt{aspin2003,kospal2015,kospal2016b}), or they may even be close binaries (like FU~Ori or L1551~IRS~5, \citealt{wang2004,rodriguez1998}). Multiplicity affects disk formation and evolution \citep{tobin2018}. In the SMA sample of eruptive young stars, \citet{liu2018} found that binaries or triple systems are systematically fainter than the rest of the sources.

In this paper, we present new ALMA 1.33\,mm continuum observations of ten FUors and FUor-like objects, including some multiple systems. In Section \ref{sec:obs} we explain the target selection and describe the observations and data reduction. In Section \ref{sec:res} we show our ALMA images and derive basic disk properties by Gaussian fitting in the image space and analytical disk model fitting in the visibility space. We note that due to the finite sensitivity of our observations and given that ALMA filters out the emission on large spatial scales, our data mostly trace structures on the disks' spatial scale, even though some of our targets are embedded in envelopes. In Section \ref{sec:rtmodeling} we detail our radiative transfer disk modeling. We explain our new results on individual objects in the context of the literature in Section \ref{sec:notes}, we discuss the whole sample and draw our conclusions in Section \ref{sec:discussion}, and summarize our main findings in Section \ref{sec:summary}.


\section{Observations and data reduction}
\label{sec:obs}

\subsection{Target selection}
\label{sec:targetselection}

For our project, we selected those FUors from the eruptive star list of \citet{audard2014} that were not included in any previous ALMA project, but still observable from the ALMA site. These were AR~6A/6B, Bran~76, HBC~687, Haro~5a~IRS, OO~Ser, V346~Nor, V900~Mon, and L1551~IRS~5. We added two extra FUors which were discovered after the publication of the \citet{audard2014} paper, V899~Mon and V960~Mon. Basic data (distances and coordinates) for our targets are shown in \autoref{tab:log}. Recently, \citet{connelley2018} completed a near-infrared spectroscopic survey of FUors and classified them based on various criteria as bona fide FUors (whose spectra display the necessary near-IR characteristics and an eruption was observed), FUor-like objects (whose spectra display the necessary near-IR characteristics but no eruption was observed), and peculiar objects (which have or had some spectral similarities to FUors, but are different enough that a FUor classification is not warranted). Based on their classification, from our sample V900~Mon and V960~Mon are bona fide FUors, L1551~IRS~5, Haro~5a~IRS, Bran~76, and HBC~687 are FUor-like objects, while V346~Nor and AR~6A/6B are peculiar objects. V899~Mon and OO~Ser were not studied in \citet{connelley2018}.

\subsection{ALMA observations and data processing}

We observed our targets with ALMA in Band 6 between 2016 October 9 and 2018 April 5. A detailed log of observations can be found in \autoref{tab:log}. For each object we took observations with the 7\,m array and two different configurations of the 12\,m array (TM2 being the more compact C40--3 configuration and TM1 being the more extended C40--6 configuration). We used a standard frequency setup that covered the $J$=2--1 line of $^{12}$CO, $^{13}$CO, and C$^{18}$O in 0.58\,GHz wide spectral windows, and defined two 1.875\,GHz wide windows centered at 234.0\,GHz and at 216.5\,GHz to cover the emission from the thermal dust continuum. Here, we focus on the continuum results, while the molecular line data will be analyzed in subsequent papers. For the purpose of atmospheric, bandpass, flux, phase, and pointing calibration, the ALMA observatory mostly selected different bright quasars. In a few cases, the atmospheric and flux calibrator was Neptune, Titan, or Uranus.

We reduced the delivered data with the ALMA pipeline using the Common Astronomy Software Applications (CASA) package version 5.1.1 \citep{mcmullin2007}. The observational executions with high signal-to-noise ratio (SNR) were self-calibrated by using 30\,s intervals for the creation of the phase calibration tables and the total integration time for the amplitude calibration. We checked the continuum spectral windows for line emission, and selected the non-contaminated channels to extract the continuum using the \textit{uvcontsub} task in CASA. After verifying that the common baselines between the three different observing configurations provided consistent visibilities, we concatenated all observations and produced a single measurement set for each target, which recovers emission in a wide range of spatial scales up to 5$\farcs$9, the maximum recoverable scale for the C40--3 configuration in Band 6. We extracted the continuum from the combined data sets and created uniformly weighted images using the CLEAN algorithm implemented in the \textit{tclean} task for CASA (version 5.4.0). The reference frequency for all maps was fixed at 224.55\,GHz. We ran \textit{tclean} non-interactively and set the flux density threshold to three times the noise level of the dirty image. We applied primary beam correction to the final clean images. The primary beam sizes at this frequency are 28$''$ for the 12\,m antennas and 48$''$ for the 7\,m antennas. The rms of our non-primary beam corrected cleaned maps, and the sizes and position angles of our resulting beams are listed in \autoref{tab:cleanmaps}.

\begin{deluxetable*}{c@{}c@{}ccc@{}c@{}cccc}
\tablecaption{Log of our ALMA continuum observations.\label{tab:log}}
\tablewidth{0pt}
\tablehead{
\colhead{Name} & \colhead{Distance\tablenotemark{a}} & \colhead{RA$_{\rm J2000}$}	& \colhead{Dec$_{\rm J2000}$} & \colhead{Date} & \colhead{Config.\tablenotemark{b}} & \colhead{Ang. res.} & \colhead{$T_{\rm exp}$} & \colhead{PWV} & \colhead{Baselines} \\
               & \colhead{(pc)}     & \colhead{(h:m:s)}  & \colhead{($^{\circ}$:$'$:$''$)} &           &          & ($''$) & (min) & (mm) & (m)  }
\startdata
\multirow{3}{*}{L1551~IRS~5} & \multirow{3}{*}{140$\pm$10} & \multirow{3}{*}{04:31:34.08} & \multirow{3}{*}{+18:08:04.9}   & 2016 Nov 14 & 7m  & 5.2  &  4.6 & \dots & 8 --    44 \\
                             &                             &                              &                                & 2018 Apr 05 & TM2 & 0.68 &  3.6 & 0.8   & 15 --  500 \\
                             &                             &                              &                                & 2017 Jul 24 & TM1 & 0.12 & 11.3 & 0.2   & 30 -- 3637 \\
\hline
\multirow{5}{*}{Haro~5a~IRS} & \multirow{5}{*}{391$\pm$24} & \multirow{5}{*}{05:35:26.71} & \multirow{5}{*}{$-$05:03:54.9} & 2016 Oct 10 & 7m  & 5.1  &  4.1 & \dots & 8 --    48 \\
                             &                             &                              &                                & 2017 Apr 06 & 7m  & 5.1  &  4.1 & \dots & 8 --    48 \\
                             &                             &                              &                                & 2017 Aug 23 & TM2 & 0.45 &  3.1 & 2.1   & 15 --  783 \\
                             &                             &                              &                                & 2017 Jul 07 & TM1 & 0.19 &  9.3 & 0.8   & 16 -- 2647 \\
                             &                             &                              &                                & 2017 Jul 26 & TM1 & 0.11 &  9.3 & 0.4   & 16 -- 3696 \\
\hline
\multirow{5}{*}{V899~Mon}    & \multirow{5}{*}{769$^{+36}_{-33}$}   & \multirow{5}{*}{06:09:19.28} & \multirow{5}{*}{$-$06:41:55.4} & 2016 Oct 09 & 7m  & 5.1  &  4.1 & \dots & 8 --    48 \\
                             &                             &                              &                                & 2017 Apr 29 & TM2 & 0.67 &  3.1 & 2.0   & 15 --  459 \\
                             &                             &                              &                                & 2018 Mar 22 & TM2 & 0.46 &  3.1 & 2.2   & 15 --  783 \\
                             &                             &                              &                                & 2017 Jul 05 & TM1 & 0.19 &  9.3 & 0.6   & 21 -- 2647 \\
                             &                             &                              &                                & 2017 Jul 27 & TM1 & 0.11 &  9.3 & 0.5   & 16 -- 3696 \\
\hline
\multirow{3}{*}{AR~6A/6B}    & \multirow{3}{*}{719$\pm$16} & \multirow{3}{*}{06:40:59.31} & \multirow{3}{*}{+09:35:52.0}   & 2016 Oct 09 & 7m  & 5.1  &  4.1 & \dots &  8 --   48 \\
                             &                             &                              &                                & 2017 Apr 20 & TM2 & 0.70 &  3.1 & 2.3   & 16 --  459 \\
                             &                             &                              &                                & 2017 Jul 22 & TM1 & 0.13 & 10.3 & 1.0   & 30 -- 3636 \\
\hline
\multirow{3}{*}{V900~Mon}    & \multirow{3}{*}{1100$\pm$100} & \multirow{3}{*}{06:57:22.22} & \multirow{3}{*}{$-$08:23:17.6} & 2016 Oct 11 & 7m  & 4.7  &  8.1 & \dots & 8 --    47 \\
                             &                             &                              &                                & 2017 Apr 20 & TM2 & 0.70 &  6.1 & 2.2   & 16 --  459 \\
                             &                             &                              &                                & 2017 Jul 27 & TM1 & 0.11 & 18.6 & 0.5   & 16 -- 3696 \\
\hline
\multirow{3}{*}{V960~Mon}    & \multirow{3}{*}{1574$^{+173}_{-143}$}  & \multirow{3}{*}{06:59:31.58} & \multirow{3}{*}{$-$04:05:28.1} & 2016 Oct 11 & 7m  & 4.7  &  8.1 & \dots & 8 --    47 \\
                             &                             &                              &                                & 2017 Apr 20 & TM2 & 0.70 &  6.1 & 2.2   & 16 --  459 \\
                             &                             &                              &                                & 2017 Jul 27 & TM1 & 0.11 & 18.6 & 0.5   & 16 -- 3696 \\
\hline
\multirow{3}{*}{Bran~76}     & \multirow{3}{*}{1060$^{+29}_{-28}$}  & \multirow{3}{*}{07:50:35.60} & \multirow{3}{*}{$-$33:06:23.9} & 2016 Oct 21 & 7m  & 4.7  &  8.3 & \dots &  8 --   47 \\
                             &                             &                              &                                & 2017 Apr 20 & TM2 & 0.70 &  6.2 & 2.3   & 16 --  459 \\
                             &                             &                              &                                & 2017 Jul 30 & TM1 & 0.10 & 20.7 & 0.8   & 16 -- 3637 \\
\hline
\multirow{3}{*}{V346~Nor}    & \multirow{3}{*}{700$^{+600}_{-350}$}  & \multirow{3}{*}{16:32:32.19} & \multirow{3}{*}{$-$44:55:30.7} & 2017 May 15 & 7m  & 5.2  &  4.1 & \dots & 8 --    47 \\
                             &                             &                              &                                & 2018 Mar 16 & TM2 & 0.46 &  3.1 & 2.0   & 15 --  783 \\
                             &                             &                              &                                & 2017 Aug 25 & TM1 & 0.10 &  9.8 & 0.4   & 21 -- 3696 \\
\hline
\multirow{4}{*}{OO~Ser}      & \multirow{4}{*}{438$\pm$11} & \multirow{4}{*}{18:29.49.13} & \multirow{4}{*}{+01:16:20.7}   & 2017 Apr 30 & 7m  & 5.2  &  4.1 & \dots & 8 --    48 \\
                             &                             &                              &                                & 2017 May 14 & 7m  & 5.2  &  4.1 & \dots & 8 --    47 \\
                             &                             &                              &                                & 2018 Mar 27 & TM2 & 0.46 &  3.1 & 1.9   & 15 --  783 \\
                             &                             &                              &                                & 2017 Jul 22 & TM1 & 0.12 &  9.8 & 0.5   & 16 -- 3696 \\
\hline
\multirow{3}{*}{HBC~687}     & \multirow{3}{*}{400}        & \multirow{3}{*}{19:29:00.86} & \multirow{3}{*}{+09:38:42.9}   & 2016 Nov 18 & 7m  & 5.2  &  9.3 & \dots & 8 --    44 \\
                             &                             &                              &                                & 2017 Apr 30 & TM2 & 0.67 &  7.3 & 1.6   & 13 --  459 \\
                             &                             &                              &                                & 2017 Jul 22 & TM1 & 0.12 & 22.8 & 0.5   & 16 -- 3696 \\
\enddata
\tablenotetext{a}{References for the distances are: L1551~IRS~5: \citet{kenyon1994c}, Haro~5a~IRS: \citet{grossschedl2018}, V899~Mon: \citet{bailerjones2018}, AR~6A/6B: \citet{maizapellaniz2019}, V900~Mon: \citet{reipurth2012}, V960~Mon: \citet{bailerjones2018}, Bran~76: \citet{bailerjones2018}, V346~Nor: \citet{graham1985}, OO~Ser: \citet{herczeg2019}, HBC~687: \citet{kospal2008}.}
\tablenotetext{b}{TM1 means the C40-6 configuration of the 12\,m array, while TM2 means the C40-3 configuration.}
\end{deluxetable*}

\begin{deluxetable}{cccc}
\tablecaption{CLEAN beam and noise.\label{tab:cleanmaps}}
\tablewidth{0pt}
\tablehead{
\colhead{Target} & \colhead{Beam size} & \colhead{Beam PA} & \colhead{rms} \\
& \colhead{($''$)} & \colhead{($^{\circ}$)}  & \colhead{(mJy beam$^{-1}$)}}
\startdata
L1551~IRS~5 & 0.14$\times$0.10 & 85.7  & 0.374 \\
Haro~5a~IRS & 0.18$\times$0.18 & $-$85.5 & 0.138 \\
V899~Mon    & 0.15$\times$0.15 & 84.8  & 0.023\\
AR~6A/6B    & 0.17$\times$0.17 & $-$87.9 & 0.046\\
V900~Mon    & 0.13$\times$0.13 & 85.6  & 0.030\\
V960~Mon    & 0.14$\times$0.14 & 85.5  & 0.028\\
Bran~76     & 0.13$\times$0.13 & $-$77.5 & 0.027\\
V346~Nor    & 0.13$\times$0.13 & 82.1  & 0.031\\
OO~Ser      & 0.15$\times$0.15 & 85.7  & 0.064\\
HBC~687     & 0.14$\times$0.14 & 87.8  & 0.042\\
\enddata
\end{deluxetable}


\section{Results}
\label{sec:res}

\begin{figure*}
\centering
\includegraphics[height=0.9\textheight]{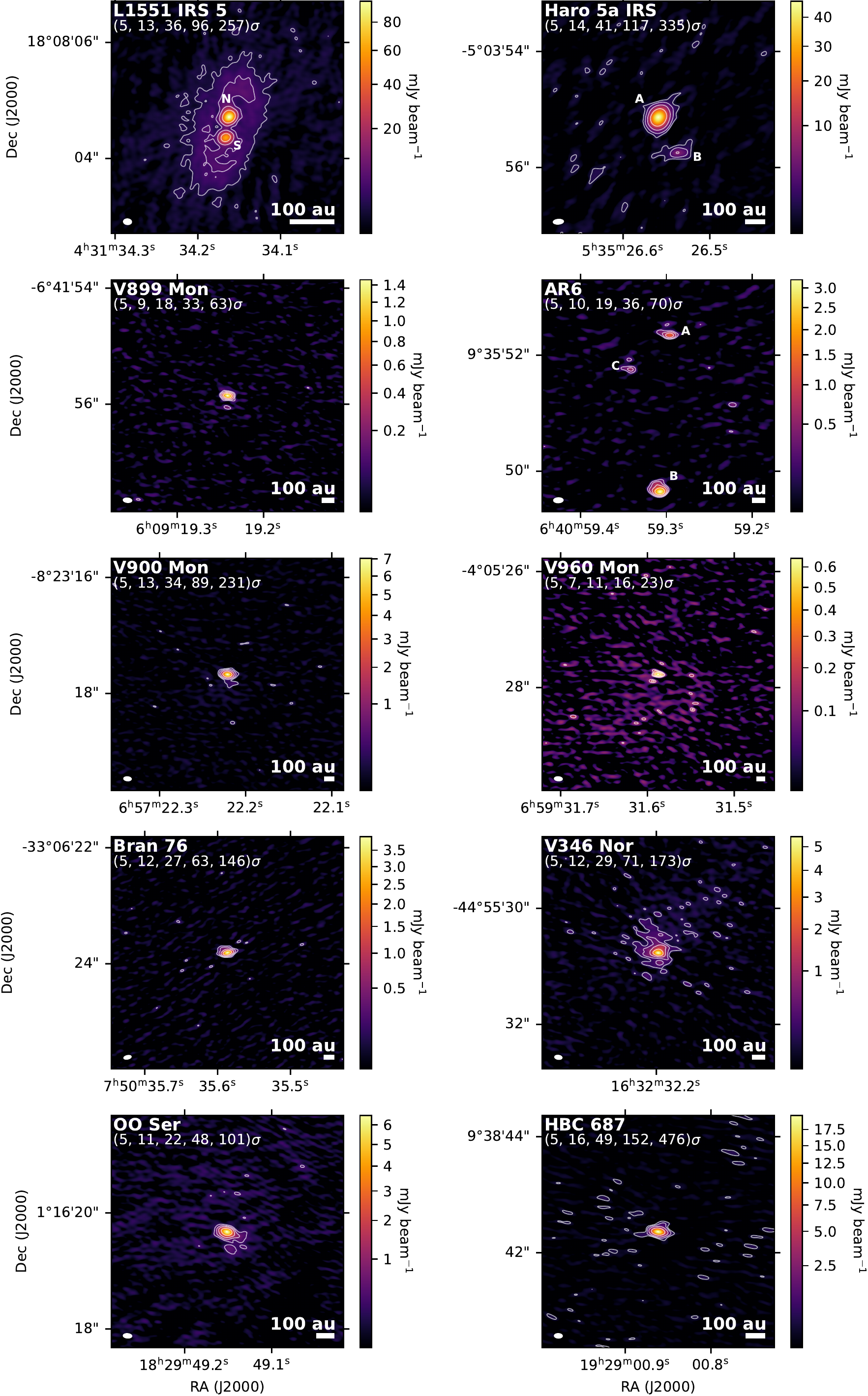}
\caption{1.33\,mm cleaned non-primary beam corrected ALMA images of our sample. Each panel is 4$''\times$4$''$. The synthesized beam is indicated in the lower left corner of each panel. The target name and the contour levels are indicated in the upper left corner of the panels. The beam sizes and rms noise ($\sigma$) values are listed in \autoref{tab:cleanmaps}.
\label{fig:alma1}}
\end{figure*}

\autoref{fig:alma1} shows our 1.33\,mm images corresponding to the thermal dust continuum emission. All of our targets were detected with high peak SNR of at least 23. Some of our maps show multiple sources. In the map of the AR~6A/6B system, we detect three sources that can be identified as 6A, 6B, and 6C \citep[cf.][]{aspin2003}. In addition, we also detected a bright source to the southwest of AR~6B, which we have labeled AR~6D and which is outside of the area plotted in \autoref{fig:alma1}. In the map of Haro~5a~IRS, there is a faint source near our target, which had not been previously reported in the literature. In the following, we will call this source Haro~5a~IRS~B. Around V960~Mon, there are three nearby sources, two of which can be identified as V960~Mon~N and V960~Mon~SE \citep[cf.][]{kospal2015}. In the following, we will call the third one V960~Mon~E. V370~Ser was detected in our OO Ser map at a separation of $\sim$11$\farcs$5 to the north of the FUor (not visible in \autoref{fig:alma1} as it falls outside of the plotted region). We detected the continuum source to the northwest of V346~Nor which had previously been reported by \citet{kospal2017c}. At a separation of $15\farcs2$, this source also falls outside of the region plotted in \autoref{fig:alma1}.

The brightness distribution of our targets typically has a centrally peaked, bright, rather compact central part, in some cases surrounded by  extended faint emission. To derive basic parameters such as coordinates, sizes, inclinations, position angles, and fluxes, we used CASA's IMFIT tool to find best-fit 2D Gaussian models for each FUor disk. The radii were estimated from the deconvolved best-fitted 2D Gaussian, we took the FWHM of the major axis and divided it by $2\sqrt{2\ln{2}}$ to obtain $\sigma$, which we use as a proxy for the radius of the disk based on their primary beam corrected images. We also estimated their masses by assuming optically thin emission, a dust opacity of $\kappa_\nu$, dust temperature of $T_{\rm dust}$, and using the following equation:
\begin{equation} \label{eq:1}
    M_{\rm dust} = \frac{F_{\nu}d^2}{\kappa_{\nu}B_{\nu}(T_{\rm dust})}.
\end{equation}
The dust opacity was estimated following \citet{beckwith1990}, where $\kappa_\nu = 0.1(\nu/10^{12}\mathrm{Hz})^\beta$, assuming an opacity power-law index of $\beta$ = 1, which resulted in $\kappa_\nu$=2.254\,cm$^2$\,g$^{-1}$ at 1.33\,mm.

The dust temperature was calculated following Equation 2 of \citet{Tobin2020}:
\begin{equation}
T_\mathrm{bol}^\mathrm{dust} = 43\,\mathrm{K}\,\left(\dfrac{L_\mathrm{bol}}{1\,L_\odot}\right)^{0.25}
\end{equation}
In the case of the known FUors, we used the bolometric luminosities measured during their outburst phase \citep[and references therein]{audard2014}. For the non-outbursting sources in our field of views we used the median luminosity estimated from 230 protostars \citep{dunham2013}, $L_\mathrm{bol}$ = 1.3 $L_\mathrm{\odot}$, which resulted in $\sim$47\,K and we rounded to $T_\mathrm{bol}$ = 50\,K. We note that other works have used lower temperatures, e.g. 20\,K \citep{ansdell2016,ansdell2017}, which would result in higher mass estimates.

We then converted the obtained dust masses to total masses assuming the usual gas-to-dust mass ratio of 100. We emphasize that the millimeter emission is likely (partially) optically thick, so these values should be considered as lower limits for the true mass \citep[cf.][]{dunham2014b,liu2019,zhu2019}. The obtained parameters are shown in \autoref{tab:gaussian}. With the 2D Gaussian fitting procedure, V899~Mon, AR~6A, and AR~6C appear to be unresolved, therefore we could not determine radius, inclination, and position angle for them. However, we provide upper limits for their radii.

We measured the in-band spectral index $\alpha$ of our targets by constructing maps from the two continuum spectral windows separately, at \mbox{$\nu_1$ = 216.9 GHz} and at \mbox{$\nu_2$ = 232.2 GHz}, and measuring the targets' fluxes ($F$) at these two frequencies. Then we calculated $\alpha$ using:
\begin{equation}
\alpha = \frac{\log{F_{\nu_1}} - \log{F_{\nu_2}}}{\log{\nu_1}-\log{\nu_2}},
\end{equation}
and its uncertainties ($\sigma$) with:
\begin{equation}
\sigma_\alpha^2 = \left(\frac{1}{\ln{\nu_1}-\ln{\nu_2}}\right)^2 \left(\frac{\sigma_{\nu_1}^2}{F_{\nu_1}^2}+\frac{\sigma_{\nu_2}^2}{F_{\nu_2}^2}\right).
\end{equation}
If the dust emission comes from large grains that emit as blackbodies or if the emission is optically thick, $\alpha$ is expected to be $\alpha = 2$ \citep{pavlyuchenkov2019}. If the emission is optically thin and the grain size distribution is typical for the interstellar matter, then the millimeter spectral slope is expected to be steeper, with $\alpha \approx 3.7$ \citep{draine2006}. The measured indices are presented in \autoref{tab:gaussian}. We obtained spectral indices between 0.27 and 4.95, however these values are highly uncertain due to the small difference between frequencies $\nu_1$ and $\nu_2$. We did not make further analysis on these values.

\begin{deluxetable*}{cR@{${}\pm{}$}LR@{${}\pm{}$}LR@{${}\pm{}$}LR@{${}\pm{}$}LccccR@{${}\pm{}$}L}
\tablecaption{Basic parameters of the detected sources from Gaussian fits.\label{tab:gaussian}}
\tablewidth{0pt}
\tablehead{
\colhead{Name} & \multicolumn2c{Radius} & \multicolumn2c{$i$} & \multicolumn2c{PA} & \multicolumn2c{Int. Flux} & \colhead{Mass (50\,K)} & \colhead{$L_\mathrm{bol}$} & \colhead{$T_\mathrm{bol}^\mathrm{dust}$} & \colhead{Mass ($T_\mathrm{bol}^\mathrm{dust}$)} & \multicolumn2c{$\alpha$} \\
& \multicolumn2c{(au)} & \multicolumn2c{($^{\circ}$)}  & \multicolumn2c{($^{\circ}$)} & \multicolumn2c{(mJy)} & ($M_{\odot}$) & ($L_{\odot}$) & (K) & ($M_{\odot}$) & \multicolumn2c{}}
\startdata
\multicolumn{15}{c}{FUors} \\
\hline
L1551~IRS~5~N & 10.6  & 0.7    & 44    & 7     & 158   & 12    & 248.7 & 9.4   & \sci{1.48}{-2} & 30\tablenotemark{a}    & 101  & \sci{6.96}{-3} & 2.08 & 0.93 \\
Haro~5a~IRS   & 39    & 2      & 55    & 1     & 169   & 2     & 166.4 & 1.7   & \sci{7.73}{-2} & 50\tablenotemark{b}    & 114  & \sci{3.18}{-2} & 1.41 & 0.36 \\
V899~Mon      & \multicolumn{2}{l}{$<$37} & \multicolumn{2}{l}{\dots} & \multicolumn{2}{l}{\dots} & 1.6  & 0.1 & \sci{2.88}{-3} & 150\tablenotemark{c} & 151 & \sci{8.87}{-4} & 3.78 & 1.15 \\
AR~6A         & \multicolumn{2}{l}{$<$38} & \multicolumn{2}{l}{\dots} & \multicolumn{2}{l}{\dots} & 1.5  & 0.2 & \sci{2.36}{-3} & 450\tablenotemark{d} & 198 & \sci{5.48}{-4} & 4.95 & 2.89 \\
AR~6B         & 31    & 4      & 54    & 27    & 12    & 27    & 4.9   & 0.4   & \sci{7.70}{-3} & 450\tablenotemark{d}   & 198  & \sci{1.79}{-3} & 2.23 & 2.01 \\
V900~Mon      & 20    & 2      & 28    & 20    & 169   & 73    & 8.4   & 0.1   & \sci{3.09}{-2} & 106\tablenotemark{e}   & 138  & \sci{1.04}{-2} & 2.62 & 0.51 \\
V960~Mon      & 37    & 28     & 61    & 28    & 71    & 20    & 0.7   & 0.1   & \sci{5.27}{-3} & \dots & \dots & \dots  & 3.51 & 4.11\\
Bran~76       & 24    & 3      & 45    & 11    & 79    & 18    & 4.8   & 0.1   & \sci{1.64}{-2} & 287\tablenotemark{f}   & 177  & \sci{4.28}{-3} & 2.13 & 0.75 \\
V346~Nor      & 42    & 28     & 38    & 10    & 105   & 23    & 11.7  & 0.7   & \sci{1.74}{-2} & 160\tablenotemark{g}   & 153  & \sci{5.29}{-3} & 3.44 & 1.92 \\
OO~Ser        & 26    & 2      & 50    & 6     & 60    & 10    & 12.8  & 0.5   & \sci{7.46}{-3} & 36\tablenotemark{h}    & 102  & \sci{3.48}{-3} & 2.79 & 1.38 \\
HBC~687       & 20    & 2      & 64    & 1     & 77    & 1     & 30.5  & 0.4   & \sci{1.48}{-2} & 10\tablenotemark{i}    & 77   & \sci{9.33}{-3} & 2.40 & 0.31 \\
\hline
\multicolumn{15}{c}{Neighboring sources} \\
\hline
L1551~IRS~5~S & 9.5   & 1.1    & 25    & 23    & 160   & 77    & 153.2 & 9.3   & \sci{9.12}{-3} & 10\tablenotemark{a}    & 77   & \sci{5.74}{-3} & 3.52 & 1.6 \\
Haro~5a~IRS~B & 34    & 6      & 56    & 7     & 97    & 10    & 13.4  & 1.3   & \sci{6.22}{-3} & \dots & \dots & \dots  & 0.27 & 2.26\\
AR~6C         & \multicolumn{2}{l}{$<$38} & \multicolumn{2}{l}{\dots} & \multicolumn{2}{l}{\dots} & 0.7  & 0.1   & \sci{1.10}{-3} & \dots & \dots  & \dots & 4.77 & 3.88 \\
AR~6D         & 50    & 7      & 56    & 14    & 43    & 14    & 4.1   & 0.3   & \sci{6.44}{-3} & \dots & \dots  & \dots & 2.50 & 1.60\\
V960~Mon~N    & 63    & 16     & 34    & 41    & 33    & 57    & 1.2   & 0.1   & \sci{9.03}{-3} & \dots & \dots & \dots  & 0.95 & 1.95\\
V960~Mon~E    & 126   & 29     & 65    & 8     & 96    & 11    & 1.5   & 0.2   & \sci{1.13}{-2} & \dots & \dots & \dots  & 0.28 & 3.03\\
V960~Mon~SE   & 227   & 25     & 73    & 2     & 170   & 2     & 7.6   & 0.3   & \sci{5.72}{-2} & \dots & \dots & \dots  & 1.74 & 1.00\\
V346~Nor~B    & 84    & 57     & 67    & 4     & 145   & 4     & 17.8  & 1.1   & \sci{2.65}{-2} & \dots & \dots & \dots  & 2.69 & 1.36 \\
V370~Ser      & 57    & 6      & 55    & 8     & 15    & 9     & 5.9   & 0.6   & \sci{3.44}{-3} & \dots & \dots & \dots  & 1.23 & 1.96 \\
\enddata
\tablecomments{References for $L_\mathrm{bol}$: \emph{a}: \citet{Liseau2005}, \emph{b}: \citet{ReipurthAspin1997}, \emph{c}: \citet{ninan2015}, \emph{d}: \citet{aspin2003}, \emph{e}: \citet{reipurth2012}, \emph{f}: \citet{green2006}, \emph{g}: \citet{kospal2017b}, \emph{h}: \citet{kospal2007}, \emph{i}: \citet{kospal2008}. Disk masses were obtained from the integrated fluxes using \autoref{eq:1} and should be considered as lower limits. Disk radii are deconvolved sizes.}
\end{deluxetable*}

\subsection{Fitting in the visibility space}
\label{sec:vismodeling}

Fitting disk properties in the image space is sensitive to the deconvolution procedure used. Therefore, to obtain results independently of the deconvolution, we fitted the disk emission in the visibility space using a procedure similar to that described in \citet{Manara19}. Based on the continuum images, we assumed that the disks were axisymmetrical and modeled each of them using a radial profile:
\begin{equation}
I(r) = I_0\,r^{-\gamma_1} \exp\left(-\frac{r}{R_{\rm vis}}\right)^{\gamma_2},
\end{equation}
where $I_0$ is the peak emission of the disk in Jy\,sr$^{-1}$, $R_{\rm vis}$ is the characteristic radius of the disk in arcseconds, $\gamma_1$ is the exponent of the power law and describes the flatness of the inner parts of the brightness radial profile, and $\gamma_2$ is the exponent of the exponential cutoff which dictates the cutoff of the outer parts of the radial profile. We used the Galario\footnote{https://mtazzari.github.io/galario/index.html} code \citep{Tazzari18} to apply the appropriate inclination and position angle ($i$ and PA, respectively) and the offsets with respect to the phase center ($\Delta$RA and $\Delta$Dec), and transform the radial profile model into visibilites ($V_{\rm m}$). Therefore, each disk has been fitted with eight free parameters: $I_0$, $R_{\rm vis}$, $\gamma_1$, $\gamma_2$, $i$, PA, $\Delta$RA, and $\Delta$Dec. We explored the parameter space with a Goodman \& Weare affine invariant Markov chain Monte Carlo using the Python ensemble sampler emcee\footnote{https://emcee.readthedocs.io/en/stable/} \citep{ForemanMackey13}. Our likelihood function was also calculated with Galario following:
\begin{equation}
\begin{split}
\chi^2 = \sum_{j=1}^N \big(&[\operatorname{Re}(V_{\rm{o},j})-\operatorname{Re}(V_{\rm{m},j})]^2 +\\
        &[\operatorname{Im}(V_{\rm{o},j})-\operatorname{Im}(V_{\rm{m},j})]^2\big) w_j,
\end{split}
\end{equation}
where $N$ is the number of visibility points and $w_j$ is the weight corresponding to each point.

For the cases where our field of view include more than one source (i.e., Haro~5a~IRS, AR~6, V960~Mon, V346~Nor, OO~Ser), we used the same model for them, calculated the visibilities individually, and added them together before calculating $\chi^2$. The initial values of the exploration of the parameter space were based on the results of our 2D Gaussian fitting (\autoref{tab:gaussian}). We utilized flat priors for all variables. Depending on the number of fitted disks, we used a different number of walkers and steps. In cases where only one disk is in the field of view, we used 150 walkers and at least 10\,000 steps. In the cases where more that one disk were detected, we used 250 walkers and between 25\,000 and 35\,000 steps. For all cases, we discarded a few thousand steps as the ``burn in'' phase and built the posterior distributions with at least 5000 steps. The best-fitted values were taken from the maximum likelihood estimate of the posterior distribution for each parameter and are presented in \autoref{tab:visibilities}. The uncertainties for each parameter were derived from the 16th and 84th percentiles of their respective posterior distributions.

We extracted the sizes of the disks from the best-fitted values following \citet{Manara19}.
The disk radius ($R_{\rm disk}$) contains 95\% of the continuum flux and the effective radius ($R_{\rm eff}$) contains 68\% of the continuum flux.
The two radii are defined as follows:
\begin{equation}
\int_0^{R_{\rm disk}} 2\pi I(r) r \,\mathrm{d}r / F_{\rm tot} = 0.95,
\end{equation}
\begin{equation}
\int_0^{R_{\rm eff}} 2\pi I(r) r \,\mathrm{d}r / F_{\rm tot} = 0.68,
\end{equation}
where $I(r)$ was calculated using the best-fitted parameters and $F_{\rm tot}$ is the total flux of the continuum emission, defined from:
\begin{equation}
F_{\rm tot}  = I_0 \cos{i} \int_0^\infty 2\pi r r^{-\gamma_1} \exp\left(-\frac{r}{R_{vis}}\right)^{-\gamma_2}\,\mathrm{d}r.
\end{equation}
Finally, we corrected $F_{\rm tot}$ based on the inclination of the disk. The calculated radii and $F_{\rm tot}$ values are also presented in \autoref{tab:visibilities}.

In the case of barely resolved disks, degeneracies would cause the MCMC to not converge into a solution. Hence, for the three sources unresolved in the Gaussian fitting (i.e. V899~Mon, AR~6A and AR~6C), we fixed the exponents to $\gamma_1=0$ and $\gamma_2=50$ to describe those disks with a flat surface profile whose emission decreases steeply after a certain radius. We examined the effects of this decision by fitting one of the well resolved sources (Bran~76) with the exponents fixed to these values, and compared our results to those obtained with the run where all parameters were free. We found that $i$, PA and $F_\mathrm{tot}$ are the same within their uncertainties, and $R_\mathrm{eff}$ and $R_\mathrm{disk}$ are smaller for the run with the fixed exponents. Therefore, we expect that in the case of the unresolved sources, the true disk radius lies between the $R_\mathrm{eff}$ from the visibility fitting and the upper limit estimated from the Gaussian fitting.

In the case of Haro~5a~IRS the fitting would not converge into a solution. To overcome this, we fixed both angles ($i$ and PA) for the two disks to the values obtained in the Gaussian fitting and obtained a solution.

The cleaned images for the data, model, and residuals, as well as the observed and model visibilities as function of $uv$ distance are plotted in \autoref{sec:app_a}.

{\renewcommand{\arraystretch}{1.2}
\begin{deluxetable*}{cCCCCCCCCC}[!ht]
\tablecaption{Results of visibility fitting\label{tab:visibilities}}
\tablewidth{0pt}
\tablehead{\colhead{Name} & \colhead{$\log_{10} I_0$} & \colhead{$R_{\rm vis}$}    & \colhead{$i$} & \colhead{PA}    & \colhead{$\gamma_1$} & \colhead{$\gamma_2$} & \colhead{$F_{\rm tot}$} & \colhead{$R_{\rm eff}$} & \colhead{$R_{\rm disk}$}\\
                          & \colhead{(Jy/sr)}         & \colhead{($''$)} & \colhead{($^{\circ}$)}       & \colhead{($^{\circ}$)} &                      &                      & \colhead{(mJy)}         & \colhead{(au)}             & \colhead{(au)}}
\decimals
\startdata
\multicolumn{10}{c}{FUors} \\
\hline
Haro~5a~IRS & 17.17_{-0.11}^{+0.10} & 3.05_{-0.38}^{+0.27} & 55 & 169 & -0.86_{-0.02}^{+0.02} & 54.86_{-6.80}^{+4.92} & 160.65_{-0.15}^{+0.17} & 33_{-2}^{+2} & 38_{-2}^{+2} \\
V899~Mon & 10.86_{-0.19}^{+0.16} & 0.90_{-0.21}^{+0.25} & 58_{-38}^{+23} & 44_{-21}^{+38} & 0 & 50 & 1.59_{-0.50}^{+0.10} & 16_{-4}^{+5} & 19_{-5}^{+5} \\
AR~6A & 11.06_{-0.27}^{+0.33} & 0.69_{-0.26}^{+0.30} & 69_{-27}^{+16} & 67_{-37}^{+47} &  0 & 50 & 1.27_{-0.86}^{+0.25} & 12_{-4}^{+5} & 14_{-5}^{+6} \\
AR~6B & 14.13_{-3.24}^{+3.56} & 4.36_{-1.34}^{+2.31} & 72_{-6}^{+4} & 7_{-4}^{+4} & -0.52_{-0.53}^{+0.47} & 106.93_{-41.26}^{+48.88} & 4.87_{-0.15}^{+0.14} & 43_{-3}^{+4} & 49_{-4}^{+5} \\
V900~Mon & 9.12_{-5.53}^{+6.42} & 0.64_{-0.24}^{+0.29} & 34_{-16}^{+15} & 181_{-21}^{+27} & 0.34_{-0.93}^{+0.71} & 45.33_{-19.90}^{+39.52} & 7.24_{-0.16}^{+0.52} & 15_{-2}^{+2} & 18_{-3}^{+5} \\
V960~Mon & 14.95_{-8.50}^{+9.85} & 0.83_{-0.36}^{+0.79} & 58_{-23}^{+15} & 88_{-25}^{+41} & -0.59_{-1.52}^{+1.18} & 63.19_{-19.93}^{+18.35} & 0.74_{-0.13}^{+0.14} & 30_{-15}^{+22} & 35_{-16}^{+25} \\
Bran~76 & -0.04_{-1.69}^{+3.20} & 1.75_{-1.01}^{+1.94} & 52_{-11}^{+9} & 52_{-10}^{+12} & 1.53_{-0.43}^{+0.23} & 29.11_{-20.16}^{+37.64} & 5.84_{-0.70}^{+2.13} & 23_{-7}^{+5} & 46_{-14}^{+18} \\
V346~Nor & -0.39_{-0.09}^{+0.10} & 4.24_{-3.17}^{+12.57} & 25_{-7}^{+5} & 2_{-2}^{+5} & 1.57_{-0.01}^{+0.01} & 0.30_{-0.24}^{+0.76} & 46.90_{-17.77}^{+56.17} & 707_{-567}^{+752} & 882_{-652}^{+690} \\
OO~Ser & 6.56_{-1.24}^{+1.04} & 3.12_{-1.07}^{+1.90} & 31_{-7}^{+4} & 32_{-8}^{+7} & 0.58_{-0.16}^{+0.18} & 43.30_{-15.98}^{+29.88} & 15.57_{-0.32}^{+0.46} & 30_{-3}^{+3} & 37_{-4}^{+5} \\
HBC~687 & 24.01_{-0.41}^{+0.44} & 0.35_{-0.03}^{+0.03} & 56_{-0}^{+0} & 74_{-0}^{+0} & -1.77_{-0.06}^{+0.06} & 20.58_{-1.59}^{+1.62} & 31.19_{-0.03}^{+0.03} & 13_{-1}^{+1} & 14_{-1}^{+1} \\
\hline
\multicolumn{10}{c}{Neighboring sources} \\
\hline
Haro~5a~IRS~B & 16.99_{-0.27}^{+0.17} & 2.17_{-0.45}^{+0.52} & 56 & 97 & -1.04_{-0.03}^{+0.04} & 32.29_{-6.59}^{+7.76} & 12.32_{-0.16}^{+0.16} & 42_{-3}^{+3} & 47_{-3}^{+3} \\
AR~6C & 10.33_{-0.43}^{+0.63} & 0.94_{-0.50}^{+0.56} & 39_{-26}^{+29} & 117_{-60}^{+44} &  0 & 50 & 0.72_{-0.17}^{+0.11} & 16_{-8}^{+9} & 19_{-10}^{+11} \\
AR~6D & 40.10_{-2.51}^{+1.13} & 5.97_{-2.48}^{+3.37} & 84_{-2}^{+3} & 44_{-2}^{+1} & -4.77_{-0.17}^{+0.40} & 126.97_{-56.13}^{+74.33} & 3.94_{-0.18}^{+0.22} & 104_{-5}^{+5} & 109_{-5}^{+5} \\
V960~Mon~N & 11.11_{-9.21}^{+14.32} & 4.20_{-1.41}^{+2.75} & 41_{-10}^{+9} & 22_{-13}^{+13} & -0.79_{-2.11}^{+1.63} & 38.13_{-18.66}^{+24.41} & 1.19_{-1.19}^{+0.35} & 237_{-70}^{+178} & 273_{-82}^{+255} \\
V960~Mon~E & 27.32_{-6.67}^{+5.83} & 4.69_{-1.97}^{+2.15} & 60_{-13}^{+10} & 70_{-11}^{+10} & -3.48_{-1.20}^{+1.21} & 72.76_{-34.27}^{+33.67} & 2.04_{-2.04}^{+0.39} & 239_{-53}^{+124} & 255_{-58}^{+134} \\
V960~Mon~SE & 37.50_{-9.37}^{+2.55} & 6.03_{-0.46}^{+0.78} & 80_{-1}^{+1} & 152_{-1}^{+1} & -4.53_{-0.41}^{+1.51} & 84.03_{-16.34}^{+9.63} & 6.62_{-0.24}^{+0.28} & 342_{-38}^{+47} & 361_{-40}^{+52} \\
V346~Nor~B & 38.95_{-0.42}^{+0.15} & 13.63_{-2.24}^{+12.81} & 85_{-0}^{+0} & 150_{-0}^{+0} & -4.97_{-0.02}^{+0.07} & 115.13_{-18.18}^{+99.19} & 10.39_{-10.39}^{+0.34} & 290_{-212}^{+286} & 307_{-222}^{+303} \\
V370~Ser & 1.75_{-9.90}^{+4.86} & 4.13_{-2.37}^{+2.62} & 81_{-9}^{+6} & 17_{-9}^{+17} & 0.70_{-3.60}^{+0.65} & 33.69_{-22.23}^{+16.25} & 1.11_{-1.11}^{+0.71} & 52_{-27}^{+136} & 69_{-38}^{+186} \\
\enddata
\tablecomments{The indicated parameters are the most probable values, while the  uncertainties for each parameter were derived from the 16th and 84th percentiles of their respective posterior distributions.}
\end{deluxetable*}
}

\section{Radiative transfer modeling}\label{sec:rtmodeling}

To constrain the disk parameters (especially the mass) of each system more accurately than with a simple Gaussian or analytical disk model fitting, we adopted a radiative transfer (RT) model fitting approach similar to \citet{cieza2018}. We use the RT code {\scriptsize RADMC-3D 0.41} \citep{dullemond2012radmc} with the Python interface radmc3dPy\footnote{\url{https://www.ita.uni-heidelberg.de/~dullemond/software/radmc-3d/manual_rmcpy/index.html}} to set the code input parameters for a given model. As the luminosity of the central heating source in young outbursting systems may not be well represented by the stellar photosphere temperature, we use the same approach as \citet{cieza2018} and set the stellar blackbody temperature to 10\,000\,K and allow the stellar radius, $R_{*}$, and therefore the luminosity, to be a free parameter in the model fitting procedure\footnote{For AR~6C, L1551~IRS~5~N, V899~Mon, V900~Mon, and V960~Mon (all components except S), we found that using a stellar temperature of 5000\,K allowed for a better convergence on a best fit model.}. This will allow us to directly compare our results with those that \citet{cieza2018} obtained for V883~Ori, HBC~494, V1647~Ori, and V2775~Ori.

The density profile of a trial protoplanetary disk model is given by:
\begin{equation}
\rho =\frac{\Sigma(r,\phi)}{H_p\sqrt{(2\pi)}}\exp{\left(-\frac{z^2}{2H_p^2}\right)},
\end{equation}
where $\Sigma$ is the surface density profile, and $H_{p}$ is the pressure scale height. The surface density profile includes a power-law inner disk and an exponential out tapering \citep{andrews2009protoplanetary}:
\begin{equation}
\label{eqn:sigma}
\Sigma(r) = \Sigma_c\left(\frac{r}{R_c}\right)^\gamma \exp{\left\{-\left(\frac{r}{R_c} \right)^{2+\gamma}\right\}},
\end{equation}
where $R_{c}$ is the characteristic radius of the disk, $\gamma$ is the power-law exponent of the radial surface density profile, and $\Sigma_{c}$ is a surface density a normalization factor with
\begin{equation}
    \Sigma_c = (2-\gamma) \frac{M_{\rm disk}}{2 \pi R_{c}^{2}}.
\end{equation}
The pressure scale height is defined as:
\begin{equation}
H_p = h_c \left( \frac{r}{R_c} \right) ^{\psi}, 
\end{equation}
where $h_c$ is the ratio of the pressure scale height over radius at $R_c$ and $\psi$ is the degree of flaring for the disk. We used {\scriptsize OpacityTool}\footnote{The OpacityTool Software was obtained from https://dianaproject.wp.st-andrews.ac.uk/data-results-downloads/fortran-package/} \citep{toon81, woitke16} to generate the dust absorption and scattering parameters. We assumed isotropic scattering. We include $0.1 - 3000~\mu$m silicates and a size distribution of $-$3.5 as input to calculate dust the opacities with a volume mixture of 60\% amorphous silicates \citep[e.g.,][]{dorschner95}, 15\% amorphous carbon \citep[e.g.][]{zubko96}, and a 25\% porosity. We further assume that a canonical dust-to-gas mass ratio of 0.01.

To converge on the best fit model, we use a Metropolis-Hastings Markov chain Monte Carlo (MCMC) model fitting approach. The free parameters considered in the modelling are: stellar radius, $R_{*}$, the total disk mass, $M_{\rm disk}$, degree of disk flaring, $\psi$, characteristic radius, $R_{c}$, power law exponent of the surface density profile, $\gamma$, scale height ratio, $h_{c}$, disk inclination, $inc$, and the disk position angle, $PA$. Given the detailed calculations in {\scriptsize RADMC-3D}, a given trial model can become relatively computationally expensive depending on the trial disk parameters. Therefore we perform the MCMC modelling in the image plane as we can more quickly converge the the most probable values of the free parameters \citep[see, e.g., ][]{booth2016,white2016}. After a trial model is computed in {\scriptsize RADMC-3D}, a 1.3 mm continuum image is then produced and projected to a trial inclination and position angle. The trial model is then attenuated by the primary beam and convolved with the synthetic beam for a given observational setup. To assess the likelihood of a given model, a $\chi^{2}$ is calculated as
\begin{equation}
    \chi^{2} = \frac{(Data - Model)^{2}}{\sigma^{2}},
\end{equation}
where $\sigma$ is the observed $\sigma_{rms}$ for a given observation multiplied by the synthetic beam size in pixels \citep[see ][]{booth2016}. A trial model is accepted if a random number drawn from a uniform distribution [0,1] is less than $\alpha$, where 
\begin{equation}
    \alpha = min(e^{\frac{1}{2}(\chi^{2}_{i} - \chi^{2}_{i+1})},1).
\end{equation}

For each system, we ran 10 chains with 1000 links each (minus 100 for burn in). The most probable values and 68\% Credible Regions are summarized in \autoref{tab:radmc_fit2}. If the posterior distributions were Gaussian, these would correspond to the 1$\sigma$ uncertainties. The data minus model residual plots for each system are in Appendix \ref{sec:app_b}. We fit all FUors in our sample and most of the nearby sources detected in our images with the exception of Haro 5a~IRS~B, AR~6C, AR~6D, and V370~Ser due to their low signal-to-noise ratio. In Appendix \ref{sec:app_c} we show some examples for the posterior distributions.

\begin{deluxetable*}{ccccccccc}
\centering 
\tablecaption{Results of RADMC-3D fitting\label{tab:radmc_fit2}}
\tablewidth{0pt}
\tablehead{
Name & $R_{*}$       & $M_{\rm disk}$ & $\psi$ & $R_{c}$ & $\gamma$ & $h_{c}$   & $i$          & PA           \\
     & (R$_{\odot}$) & (M$_{\odot}$)  &        & (au)    &          & (R$_{c}$) & ($^{\circ}$) & ($^{\circ}$)}
\startdata
\multicolumn{9}{c}{FUors}\\
\hline
L1551~IRS~5~N & 3.57          & 0.24           & 1.99           & 16         & $-$0.00038           & 0.97          & 79       & 62\\
              & [3.38, 5.05]  & [0.16, 0.28]   & [1.61, 1.99]   & [15, 18]   & [$-$0.39, $-$0.00018] & [0.82, 0.98]  & [69, 83] & [56, 108]\\
Haro~5a~IRS   & 7.35          & 0.38           & 0.52           & 69         & $-$0.31              & 0.49          & 68       & 155\\
              & [4.36, 7.37]  & [0.36, 0.61]   & [0.25, 1.34]  & [57, 79]   & [$-$1.08, $-$0.17]  & [0.44, 0.78]  & [54, 76] & [150, 170]\\
V899~Mon      & 6.39          & 0.030          & 0.21           & 21         & $-$0.32              & 0.20          & 41       & 75\\
              & [2.42, 6.57]  & [0.018, 0.056]  & [0.21, 1.52]  & [14, 40]    & [$-$1.56, $-$0.28]  & [ 0.20, 0.77] & [ 14, 53] & [46, 115]\\
AR~6A         & 2.61          & 0.080          & 0.63           & 29         & $-$0.63              & 0.13          & 75       & 83\\
              & [2.02, 3.04]  & [0.038, 0.10]  & [0.39, 1.27]  & [27, 51]   & [$-$1.45, $-$0.29]  & [0.11, 0.71]  & [22, 77]  & [55, 90]\\
AR~6B         & 3.01          & 0.011          & 1.37           & 20         & $-$0.42              & 0.73          & 84       & 71\\
              & [1.83, 7.67]  & [0.0052, 0.024] & [0.45, 1.73]   & [14, 48]    & [$-$1.34, $-$0.23]  & [0.15, 0.81] & [14, 79]  & [26, 162]\\
V900~Mon      & 3.66          & 0.30           & 0.53           & 22         & $-$0.11              & 0.82          & 14       & 89\\
              & [3.43, 6.76]  & [0.24, 0.47]   & [0.37, 1.55]   & [18, 40]   & [$-$1.52, $-$0.04]  & [ 0.44, 0.85] & [10, 46]  & [66, 121]\\
V960~Mon      & 1.69          & 0.17           & 0.63           & 46         & $-$0.64              & 0.48          & 15       & 96\\
              & [1.68, 6.45]  & [0.066, 0.27] & [0.31, 1.60]  & [32, 89]  & [$-$1.39, $-$0.32]  & [0.21, 0.80] & [12, 69]  & [34, 139]\\
Bran~76       & 1.09          & 0.16           & 1.58           & 27         & $-$0.077             & 0.29          & 20       & 102\\
              & [1.05, 2.14]  & [0.089, 0.23]  & [0.42, 1.65]  & [19, 41]   & [$-$0.70, $-$0.068]  & [0.27, 0.73]  & [17, 68]  & [46, 119]\\
V346~Nor      & 1.95          & 0.18           & 0.30           & 49         & $-$0.15              & 0.41          & 20       & 105 \\
              & [1.59, 2.21]  & [0.15, 0.23]   & [0.098, 0.61]   & [48, 104]  & [$-$0.48, $-$0.082]   &  [0.33, 0.84] & [10, 30]  & [90, 107] \\
OO~Ser        & 2.80          & 0.025          & 0.21           & 25         & $-$0.1               & 0.78          & 74       & 62\\
              & [2.42, 5.10]  & [0.022, 0.048] & [0.20, 1.59]  & [22, 46]   & [$-$1.40, $-$0.19]  & [0.51, 0.90]  & [16, 74]  & [$-$34, 177]\\
HBC~687       & 2.61          & 0.053          & 1.36           & 17         & $-$0.21              & 0.67          & 37       & 66 \\
              & [2.48, 6.34]  & [0.040, 0.063] & [0.53, 1.69]   & [13, 24]    & [$-$1.48, $-$0.20]  & [0.32, 0.80]  & [21, 70]  & [40, 130]\\
\hline
\multicolumn{9}{c}{Neighboring sources}\\
\hline
L1551~IRS~5~S & 3.58          & 0.028          & 0.32           & 11         & $-$1.26              & 0.16          & 44       & 121\\
              & [2.11, 6.74]  & [0.012, 0.039] & [0.30, 1.62]  & [9, 20]    & [$-$1.64, $-$0.33]  & [0.15, 0.79] & [13, 50]  & [64, 122]\\
V960~Mon~N    & 5.24          & 0.13           & 1.37           & 77         & $-$0.42              & 0.63          & 60       & 12\\
              & [1.84, 6.65]  & [0.081, 0.30]  & [ 0.51, 1.69] & [50, 156]  & [$-$1.52, $-$0.32]  & [0.19, 0.80] & [18, 74]  & [2, 140]\\
V960~Mon~S    & 7.20          & 0.29           & 1.52           & 292        & $-$1.65              & 0.29          & 54       & 168\\
              & [4.25, 12.73] & [0.20, 0.52]   & [0.85, 1.67]   & [132, 318] & [$-$1.78, $-$1.18]   & [0.23, 0.62] & [44, 68] & [165, 183]\\
V960~Mon~E    & 2.09          & 0.19           & 1.26           & 74         & $-$0.63              & 0.70          & 19       & 91\\
              & [1.32, 6.63]  & [0.11, 0.28]  & [0.38, 1.69]  & [58, 76]   & [$-$1.49, $-$0.32]  & [0.124, 0.87]  & [12, 66]  & [73, 110]\\
V346~Nor~B    & 2.85          & 0.55           & 0.62           & 145        & $-$1.05              & 0.48          & 66       & 142 \\
              & [2.76, 6.52]  & [0.33, 0.77]   & [0.26, 1.16]  & [136, 163] & [$-$1.32, $-$0.45]   & [0.34, 0.79]  & [57, 74] & [134, 154]\\ 
\enddata
\tablecomments{The indicated parameters are the most probable values, while the ranges in the square brackets show the 68\% Credible Region for each parameter.}
\end{deluxetable*}


\section{Results on individual objects}
\label{sec:notes}

In the following, we discuss each of our targets individually. We first summarize earlier results in relation to their circumstellar properties, then describe our new ALMA results in context of the literature. To compare the results of the different types of modeling presented in this paper, we refer to \autoref{sec:app_d}. The geometrical parameters from the different models are generally in good agreement within the uncertainties (\autoref{fig:comp_plots}). However, there are significant differences between the masses obtained from the RT modeling (\autoref{tab:radmc_fit2}) and the estimates that simply convert the flux to mass (\autoref{tab:gaussian}).

To understand and quantify the differences, we calculated the mass that is within the optically thick part ($\tau>1.0$) of the disk from the best-fitting RT models for each of our targets, along with the optically thick mass fraction and the midplane temperature at the characteristic radius (\autoref{tab:massfraction}). RT masses for the FUor sample are a factor of 1.4--34 higher than the optically thin masses, i.e., masses obtained from the fluxes assuming 50\,K, and a factor of 5.7--146 higher when using the temperatures from the bolometric luminosities While there is no one-on-one correspondence between the optically thick mass fraction and the ratio of the RT masses to optically thin masses, they do generally correlate: the disks with highest/lowest optically thick mass fraction have the highest/lowest ratio between the RT and optically thin masses. These indicate that (at least part of) the disks are optically thick at 1.33\,mm, and the RT masses are higher because the raytracing takes into account optical depth effects. Therefore, the RT masses are probably closer to the real disk masses, and in the following, we will use them in the discussion.

\begin{deluxetable}{ccccc}
\tablecaption{Mass fraction of the RT generated disk that is optically thick ($\tau_{1.3\rm{mm}}>1.0$) and midplane temperature at the characteristic radius.\label{tab:massfraction}}
\tablehead{
\colhead{Name} & \colhead{$M_{\rm disk}$} & \colhead{$M_{\tau>1.0}$} & \colhead{$\frac{M_{\tau>1.0}}{M_{\rm disk}}$} & \colhead{$T_{R_C}$} \\
& \colhead{(M$_{\odot}$)} & \colhead{(M$_{\odot}$)} & & \colhead{(K)}}
\startdata
L1551 IRS 5 N & 0.24   & 0.12   & 0.48    & 130 \\  
Haro 5a IRS   & 0.38   & 0.10   & 0.27    & 125 \\  
V899 Mon      & 0.030  & 0.0093 & 0.31    & 51  \\  
AR 6A         & 0.080  & 0.041  & 0.51    & 60  \\  
AR 6B$^{*}$   & 0.011  & \dots  & $<$0.01 & 210 \\  
V900 Mon      & 0.30   & 0.19   & 0.64    & 110 \\  
V960 Mon      & 0.17   & 0.0093 & 0.055   & 33  \\  
Bran 76       & 0.16   & 0.065  & 0.41    & 82  \\  
V346 Nor      & 0.18   & 0.050  & 0.28    & 68  \\  
OO Ser$^{*}$  & 0.025  & \dots  & $<$0.01 & 220 \\  
HBC 687       & 0.053  & 0.014  & 0.26    & 240 \\
\enddata
\tablecomments{(*) These RT generated disks were completely optically thin, likely due to the combination of large scale heights and lower relative mass, and therefore we write the optically thick mass fraction as $<0.01$.}
\end{deluxetable}

\subsection*{L1551~IRS~5}
L1551~IRS~5 is a Class\,0/I FUor-like embedded object in the Taurus star-forming region. The system is a 365\,mas (51\,au) protobinary surrounded by circumstellar disks, circumbinary material \citep[and references therein]{cruzsaenzdemiera2019} and a massive envelope \citep{chou2014}. It drives a bipolar outflow \citep{Wu2009} and two jets are detected in radio \citep{rodriguez1998} and optical/near-infrared wavelengths \citep[and references therein]{Pyo2005,Fridlund2005}.

Our observations recovered the two protostellar disks and resolved the circumbinary emission. The different components of the system were already analyzed by \citet{cruzsaenzdemiera2019}, and here we will use their reported disk properties and compare them with the rest of our sample. For the Gaussian fitting, \citet{cruzsaenzdemiera2019} separated the Northern disk into two components to avoid contamination from circumbinary material, and their fitted parameters for both disks match those from 7\,mm VLA observations with higher angular resolution \citep{Lim2016}. For the radiative transfer fitting, we used a single component to model each protostellar disk and, in the case of the Northern disk, this causes differences in the inclination and position angle between the Gaussian fitting and the radiative transfer. For the total disk masses in \autoref{tab:gaussian} we made different assumptions on the disk temperature than \citet{cruzsaenzdemiera2019}. They used the brightness temperature as a proxy for the disk temperature and obtained 160\,K and 100\,K for the N and S disks, respectively, whereas here we estimated the disk temperature from the bolometric luminosity and obtained 101\,K and 77\,K, respectively.

Taking into account both our and the literature data, we can conclude that the two cirucmstellar disks have the same inclinations and positions angles, and our best estimate for them are $i=44^{\circ}$ and PA$=158^{\circ}$. These disks are the smallest in our sample. This is not surprising if we consider that this is the closest separation binary in our sample and dynamical effects might have truncated the circumstellar disks. From our radiative transfer modeling, we obtained a disk mass of 0.24$\,M_{\odot}$ for the disk of the N component. Interestingly, the S component, which is not outbursting, has a disk which is almost an order of magnitude less massive.  

\subsection*{Haro~5a~IRS}

Haro~5a~IRS is an embedded FUor-like object in the Orion Nebula Cluster, located between two sides of a bipolar nebula, Haro~5a to the west and Haro~6a to the east \citep{reipurth1997b}. Haro~5a and 6a are separated by a high extinction ridge that hides the young stellar object at optical wavelengths, but Haro~5a~IRS is well detected at longer wavelengths, e.g., in the mid- and far-infrared (IRAS 05329$-$0505, MIR~10, \citealt{nielbock2003}, HOPS~84, \citealt{fischer2010}) and in the (sub)millimeter (OMC3~MMS~7, \citealt{chini1997}, OMC3~SMM~11, \citealt{takahashi2013}).

Haro~5a~IRS is surrounded by a large amount of circumstellar material. ALMA 3\,mm continuum observations by \citet{kainulainen2017} suggest a total mass of 1.9$\,M_{\odot}$ (using 20\,K, $\kappa_{\nu} = 0.1(\nu/1000\,\rm{GHz})^{\beta}$ cm$^2$g$^{-1}$,  $\beta=1.5$, and a gas-to-dust ratio of 100). Using optically thin $^{13}$CO J=3--2 line observations, \citet{kospal2017b} obtained 1.2$\,M_{\odot}$. Considering the uncertainties in dust opacities and optical depth, these numbers are fairly consistent with each other.

In accordance to the bipolar nebula nature of the object, it is not surprising that our ALMA data shows that the disk of Haro~5a~IRS is inclined (55--68$^{\circ}$). Based on our radiative transfer model, we obtained a total disk mass of 0.38$\,M_{\odot}$, while its characteristic radius is 69\,au. Both of these parameters are the largest in our FUor sample.

We detected a nearby source at a distance of 0$\farcs$70 to the southwest (PA=209$^{\circ}$) of Haro~5a~IRS, which has not been reported earlier in the literature. We call this source Haro~5a~IRS~B. Our ALMA line data indicate CO emission from this source at approximately the same velocities as Haro~5a~IRS itself (K\'osp\'al et al.~in prep.), suggesting that they form a physical pair. Their projected separation corresponds to 273\,au. Haro~5a~IRS~B is fainter and not as well resolved as Haro~5a~IRS, therefore it must correspond to a less massive disk.

\subsection*{V899~Mon}

V899~Mon was classified as a young eruptive star by \citet{ninan2015} with  photometric and spectroscopic properties that fall between those of classical FUors and EXors. They compiled the SED of the object and modeled it using the SED fitter tool of \citet{robitaille2007} considering data points only up to 70$\,\mu$m. They found a disk mass between 10$^{-5}$ and 10$^{-3}\,M_{\odot}$ and an envelope mass between 10$^{-7}$ and 10$^{-5}\,M_{\odot}$, depending on whether they used quiescent or outburst photometry. Because the SED of V899~Mon shows a curious dip at 70$\,\mu$m and increases again until 160$\,\mu$m, \citet{ninan2015} interpreted this as a separate clump, possibly an envelope component with a large inner cavity. They obtained 20-23$\,M{_\odot}$ for the mass of this component. Following \citet{kospal2017b}, we used optically thin $^{13}$CO line observations from APEX to convert the observed line fluxes to total gas mass assuming local thermodynamic equilibrium and 20\,K temperature, integrated over an 10\,000\,au radius area. We obtained 0.24$\,M_{\odot}$, which is between the values published for the disk and the far-infrared envelope component by \citet{ninan2015}.

Our ALMA observations suggest that V899~Mon is among the less massive FUor disks in our sample, only AR~6B and OO~Ser has smaller disk masses. It has a moderate characteristic radius. Although the Gauss fitter in casaviewer found the object to be a point source, our visibility-space fitting and radiative transfer modeling could measure its size. As opposed to \citet{ninan2016}, who derived an inclination of 10$^{\circ}$, i.e., an almost face-on geometry from optical spectroscopy, our ALMA data shows an intermediate inclination (41--58$^{\circ}$). The small mass we measure with ALMA is in contrast with the massive envelope suggested by the far-infrared and submillimeter SED shape. Although our ALMA data recover all emission up to 5$\farcs$9 (or about 4500\,au), we do not detect the suspected envelope component. This may be a sensitivity issue if the envelope has a very low surface brightness, or this component may be too extended and is filtered out even with the ALMA 7m array. Indeed, \citet{ninan2015} wrote that the source's FWHM is 50$''$ in the Herschel/SPIRE 500$\,\mu$m image. Since the FWHM of SPIRE at this wavelength is only 35$\farcs$2, this hints at an envelope that extends at least about 35$''$ or several times ten thousand au.

\subsection*{AR~6A/6B}

Although no eruption was observed for them, both AR~6A and AR~6B were classified as FUors based on their spectroscopic similarities to FU~Ori and PP~13S* by \citet{aspin2003}. However, \citet{connelley2018} noticed that their spectrum of AR~6A, obtained more than ten years after that of \citet{aspin2003}, shows fewer of the spectral characteristics required for a FUor-like classification, suggesting significant spectral variability. \citet{connelley2018} did not find convincing evidence to classify AR~6B as a FUor-like object either. They classify both AR~6A and AR~6B as peculiar objects.

The objects are located in the NGC 2264 cluster, which is part of the Mon~OB1 East giant molecular cloud. \citet{forbrich2010} presented a Spitzer/IRS spectrum for AR~6A and found ice absorption features and strong PAH emission in the 5--38$\,\mu$m range. They constructed the SED of the object and tried to fit it using the SED fitter tool of \citet{robitaille2007}, but found that it was not well fitted by the models and did not give model parameters. On the other hand, \citet{gramajo2014} could fit both AR~6A and AR~6B with \citet{robitaille2007} models, and found the following disk and envelope masses: for AR~6A $M_{\rm disk}$ = 0.34$\,M_{\odot}$, $M_{\rm env}$ = 0.20$\,M_{\odot}$, for AR~6B $M_{\rm disk}$ = 0.37$\,M_{\odot}$, $M_{\rm env}$ = 0.01$\,M_{\odot}$. Using APEX CO observations,  \citet{kospal2017b} found that there is no distinct peak in the CO emission at the stellar position, concluded that the CO emission are most likely not associated with these sources, and could only give an upper limit of 1.3$\,M_{\odot}$ for the total gas mass using the optically thin $^{13}$CO(3--2) line.

We clearly detect both AR~6A and AR~6B in our ALMA images. At 1.33\,mm, AR~6B is the brightest target, then comes AR~6A, and even AR~6C is detected at a $\sim$7$\sigma$ level. The disks of AR~6A and AR~6B are not particularly large or massive, compared to the rest of our FUor sample. In fact, AR~6B has the least massive disk within our FUor sample. We measure separations of 2$\farcs$71 (A--B), 0$\farcs$89 (A--C), and 2$\farcs$18 (B--C). For comparison, \citet{aspin2003} measured a separation of 2$\farcs$8 between A--B and 0$\farcs$85 between A--C in Subaru near-infrared $JHK^{\prime}$ images. We detected an additional source 4$\farcs$43 to the southwest (PA = 244$^\circ$) of AR~6B at a $\sim$40$\sigma$ level, which we have labeled AR~6D.

\subsection*{V900~Mon}

V900~Mon went into eruption some time between 1953 and 2009 \citep{reipurth2012}. Based on  this, as well as various spectroscopic characteristics, it is classified as a bona fide FUor by \citet{connelley2018}. According to \citet{reipurth2012}, V900~Mon is a Class~I-type young stellar object embedded in a cool envelope. Its SED, if corrected for $A_V=13\,$mag extinction, is remarkably similar to that of FU~Ori. \citet{kospal2017b} measured the object with APEX in different CO lines but found no distinct CO peak towards the stellar position, suggesting a rather low-mass envelope. Therefore, an upper limit of 0.1$\,M_{\odot}$ for the total mass of the circumstellar material was given. Based on modeling the SED with \citet{robitaille2007} models, \citet{gramajo2014} found a disk mass of 0.2$\,M_{\odot}$ and an envelope mass of 0.072$\,M_{\odot}$.

Based on ALMA data, \citet{takami2019} analyzed the morphology and kinematics of the circumstellar matter around V900~Mon at $10^4$\,au spatial scale. The $^{12}$CO emission indicated an extended bipolar outflow in the east-west direction. The $^{13}$CO emission traced compressed gas in the cavity wall. The C$^{18}$O emission revealed a rotating envelope.

We clearly detected V900~Mon in our ALMA continuum image. No other millimeter sources are present in the field of view of the primary beam. With a mass of 0.3$\,M_{\odot}$ and characteristic radius of 22\,au, it has a small and massive disk. The disk is seen under a modest inclination between 14$^{\circ}$ and 34$^{\circ}$ depending on the type of model fitting. The Gaussian and visibility space fitting suggests that the orientation of the major axis of the disk is close to north-south, which means that the molecular outflow seen by \citet{takami2019} is perpendicular to the disk.

\subsection*{V960~Mon}

V960~Mon went into outburst in 2014 and shows all the necessary spectral properties of a bona fide FUor (\citealt{kospal2015} and references therein, \citealt{connelley2018}). Based on periodic oscillations in the outburst light curve, \citet{hackstein2015} suggests that it is a close binary. \citet{kospal2015}  studied the SED of the progenitor and found that it indicates the presence of a moderately flared disk and an outer envelope in the system, whose total mass is between 0.1--0.7$\,M_{\odot}$, if we scale it to the new Gaia distance of 1574\,pc from the previously assumed 450\,pc.

We detect several millimeter sources in our ALMA image. V960~Mon itself is clearly detected, just like V960~Mon~N (5\farcs53 away at PA = 15$^\circ$) and V960~Mon~SE (8\farcs45 away at PA = 152$^\circ$), two nearby sources identified previously in infrared images \citep{kospal2015}. None of the T Tauri candidates listed in \citet{kospal2015} are detected in the millimeter. One more millimeter source to the east of V960~Mon, V960~Mon~E (5\farcs36 away at PA = 97$^\circ$), has no known infrared counterpart. \citet{caratti2015} reported a companion at a separation of 227 mas and PA of 131$^{\circ}$ detected using VLT/SINFONI at near-infrared wavelengths, as well as a bump to the southwest of V960~Mon. In our ALMA image, we do not detect the bump. In the residuals of the visibility fitting for this target (\autoref{fig:vis_v960mon}), there are residuals over 5$\sigma$ at the same position angle but slightly closer (155 mas) than the companion discovered by \citet{caratti2015}. Considering that our ALMA observations were taken within two years of the SINFONI data, it is unlikely that the millimeter emission in our residual map can be associated with this companion.

Our modeling suggests that within our FUor sample, V960~Mon has a rather large disk with an average mass. The disk's position angle is close to east-west, and its inclination is between 15$^{\circ}$ and 61$^{\circ}$, depending on the type of model fitting. APEX observations of the CO emission from this source suggests an outflow oriented along the line of sight (Cruz-S\'aenz de Miera et al. in prep.), which favors the more face-on geometry for the disk.

\subsection*{Bran~76}

Bran~76 was classified as a FUor by \citet{eisloffel1990} based on high-resolution optical spectra. \citet{connelley2018} confirmed that its infrared characteristics are also FUor-like, although no outburst was ever observed. \citet{reipurth2002} noted that it has been slowly fading since 1983, although photographic plates from 1900 and 1927 suggest that the fading rate must have been lower earlier. Its SED is consistent with a steady accretion disk model shortward of 10$\,\mu$m. At longer wavelengths there is excess emission due to an envelope, whose luminosity is intermediate between that of FU~Ori and V1057~Cyg. Earlier APEX observations indicate a moderate mass envelope in the system \citep[0.02$\,M_{\odot}$,][]{kospal2017b} and no molecular outflow.

We clearly detected Bran~76 in the 1.33\,mm continuum with ALMA. Our modeling suggests that it has an average disk size and mass within the observed FUor sample. The disk has an intermediate inclination. We do not see any other source in the primary beam, but the residuals in \autoref{fig:vis_bran76} hint at some extended emission at several hundred au scale, possibly from the envelope.

\subsection*{V346~Nor}

V346~Nor was the fifth object that was discovered to undergo a FUor outburst \citep{reipurth1985}. It brightened some time between 1976 and 1980, reached peak brightness in 1992, showed a deep minimum in 2010--11, and is currently brightening again \citep{kraus2016, kospal2017a, kospal2020}. In \citet{kospal2017c} we analyzed earlier 1$''$ resolution ALMA 1.3\,mm continuum, as well as $^{12}$CO, $^{13}$CO and C$^{18}$O $J$ = 2--1 line observations of V346~Nor and found that it is surrounded by a 10\,000 au envelope with an outflow cavity. The central $\sim$700\,au part of the circumstellar matter forms a flattened pseudo-disk where material is infalling with conserved angular momentum. Within about 350 au, the velocity profile is more consistent with a disk in Keplerian rotation. The infall rate from the envelope onto the disk is 6$\,\times\,$10$^{-6}\,M_{\odot}$yr$^{-1}$, which is a factor of a few higher than the quiescent accretion rate from the disk onto the star. This hints at a mismatch between the infall and accretion rates as the cause of the eruption.

From our RT modeling we obtained a characteristic radius of 49\,au, which suggests that the detected emission is dominated by the inner part of the Keplerian disk. With its $M_{\rm disk}=0.18\,M_{\odot}$, the disk of V346~Nor is among the most massive FUor disks. This disk mass is on the same order of magnitude as the protostellar mass.

An earlier, lower spatial resolution ALMA continuum image (fig.~1 in \citealt{kospal2017c}) revealed faint extended emission around the star that seem to follow the outflow cavity walls. We detect extended emission in our new ALMA images, although with low signal-to-noise ratio, especially to the northwest of the source. The extension of this faint emission causes our single-component visibility-space fitting to flatten out (i.e., low value of $\gamma_2$) causing $R_{\rm eff}$ and $R_{\rm disk}$ to reach hundreds of au. To estimate the radius of the compact disk that can be compared with the modeling done in the image-space, we run another fit using a two-component model. This model gives $R_{\rm eff}=21$\,au and $R_{\rm disk}=25$\,au, somewhat smaller values than either the Gaussian or the RT fits. The extended component reaches out to $\sim$300 au, which matches the $\sim$0$\farcs$5 extension seen in the continuum maps (\autoref{fig:alma1} and \autoref{fig:vis_v346nor}). We also detected V346~Nor~B, a deeply embedded protostar \citet{kospal2017c} discovered in the earlier ALMA data set at a separation of 15$\farcs$2.

\subsection*{OO~Ser}

OO~Ser is a deeply embedded pre-main sequence star that went into outburst in 1995 and faded back to quiescence on a $\sim$8 years time scale \citep{hodapp1996,hodapp2012, kospal2007}. \citet{kospal2007} compiled its broad-band SED for different epochs and found data until 200$\,\mu$m, although the longer wavelength points suffer from source confusion due to nearby sources like V370~Ser, V371~Ser, EC~38, and SMM9. Deep ice absorption features in the infrared spectrum suggests the presence of a thick cold envelope \citep{kospal2007}. The optical depth of the 9.7$\,\mu$m silicate feature implies $A_V = 42\,$mag. OO~Ser is clearly visible in a JCMT/SCUBA 850$\,\mu$m image by \citet{yoo2017}, but they did not give its flux or determined its mass. \citet{kospal2017b}  detected CO line emission with APEX and determined a total circumstellar mass of 0.6$\,M_{\odot}$ using the optically thin $^{13}$CO $J=3-2$ line flux. Although the CO line profiles showed high-velocity wings, the spatial resolution of APEX was not sufficient to disentangle which sources drive molecular outflows in this region. On the other hand, \citet{hodapp1999} found H$_2$ S(1) emission to the east of OO~Ser, suggesting jet activity.

We detect two millimeter sources in our image, OO~Ser and V370~Ser (at a distance of 11$\farcs$78 to the north of OO~Ser, at PA of 7$\fdg$6). OO~Ser has a relatively low-mass, small disk within our FUor sample. It has the second smallest disk mass after AR~6B. The disk has an intermediate inclination and a position angle between 32$^{\circ}$ and 62$^{\circ}$, depending on the type of model fitting. OO~Ser has a bipolar reflection nebula \citep{hodapp1996}, extending in the east-west direction. If we interpret this as the outflow direction, and if the disk is perpendicular to that, then the scattered light morphology supports the lower position angle for the disk.

\subsection*{HBC~687}

Although it displayed no outburst, HBC~687 (also known as Parsamian~21) was first classified as a FUor-like object based on  spectroscopic properties by \citet{staude1992}, later confirmed by \citet{connelley2018}. \citet{kospal2008} obtained near-infrared polarimetric images of the scattered light nebula in the system and found a circumstellar envelope with a polar cavity and an edge-on disk with a position angle of 78$^{\circ}\pm4^{\circ}$. The disk seems to be geometrically flat and extends from approximately 48 to 360\,au from the star. The SED of HBC~687 is consistent with this picture (see also \citealt{gramajo2014}). APEX observations bt \citet{kospal2017b} revealed a relatively tenuous envelope around the system, with a total mass of $0.01\,M_{\odot}$ measured within 10\,000\,au of the central star using the optically thin $^{13}$CO(3--2) line.

With ALMA we detected a single source at the stellar position. Our modeling suggests a modest disk mass and rather small disk size within our FUor sample. While the millimeter emitting disk seems to be more inclined than perfectly edge-on, its position angle is consistent with what is seen in scattered light \citep{kospal2008}.


\section{Discussion}
\label{sec:discussion}

\subsection{Diversity of the circumstellar matter of FUors}

FUor envelopes are predicted to exhibit large diversity, depending on how far they are in the dispersal process during the FUor phase. Indeed, \citet{quanz2007c} and \citet{kospal2020b} found both FUors showing 10$\,\mu$m silicate absorption (younger, more embedded objects), and FUors with silicate emission (objects with envelopes that are already partly opened up, enabling direct view on the surface of the accretion disk). A similar evolutionary sequence was outlined by \citet{green2006} based on the amount of far-infrared excess, which traces the dust mass in the envelope. Concerning the gas component, \citet{kospal2017b} found a similar division between younger and more evolved FUors based on single-dish millimeter CO data. 

Envelopes play a significant role in the outburst physics, by both replenishing the disk material after each outburst and sustaining gravitational instability, the key ingredient of several outburst models \citep{zhu2009, vorobyov2005, vorobyov2010, vorobyov2015, kuffmeier2018}. ALMA already made an important contribution in mapping the envelope structure of a few selected FUors, resulting in the discovery of the widest outflow cavity in a Class\,I object known to date (HBC~494, \citealt{ruiz-rodriguez2017}), and the first clear detection of the transition from an infalling envelope to a rotating disk in a FUor (V346 Nor, \citealt{kospal2017c}). However, in such diverse class of objects, general conclusions cannot be drawn from data on a limited number of objects and a complete, deep, and homogeneous survey is indispensable.

FUor disks are also expected to display a large diversity while evolving from a Class\,I massive accretion disk towards the passive protoplanetary phase. While it is challenging to observationally separate disks from remnant circumstellar envelopes, earlier studies in the literature based on (sub)mm interferometry indeed revealed very different disk masses in FUors (in the range of 0.007 -- 0.3\,$M_{\odot}$, see L1551~IRS~5 in \citealt{takakuwa2004}, PP~13S* in \citealt{perez2010}, V883~Ori in \citealt{cieza2016}, HBC~494 in \citealt{ruiz-rodriguez2017}, FU~Ori in \citealt{hales2015} and \citealt{perez2020}, or V2775~Ori in \citealt{zurlo2017}). As a new example, our group is studying HBC~722, a FUor where the disk remained undetected with SMA and IRAM \citep{dunham2012, kospal2016b}, but could be detected with ALMA, obtaining a total disk mass of $0.01\,M_{\odot}$ \citep{xi2020}. This large variety in disk properties might imply different outburst mechanisms for objects in different evolutionary stages or with different disk structures.

\subsection{Differences between FUor and Class\,I/II disks}

Episodic accretion during the pre-main sequence stellar evolution may be common \citep[e.g.,][]{audard2014}. This is supported by the observational fact that accretion-related outbursts can be observed for YSOs of various evolutionary state from the very embedded to the disk only. Also, accretion outburst are observed for YSOs of various protostellar masses from $\lesssim$0.1\,$M_{\odot}$ (V346 Nor in \citealt{kospal2017c} or OO Ser in \citealt{hodapp1996}) to 20\,$M_{\odot}$ (S255IR~NIRS~3 in \citealt{carattiogaratti2017}). Therefore, it is a highly relevant question whether the disk properties of eruptive YSOs are in any way different from the disks of normal YSOs not showing signs of present or past outburst. Our current study provides an opportunity to inspect this further by comparing the disk parameters we obtained from our radiative transfer modeling of FUors to samples of regular Class\,I or Class\,II disks modeled in a similar way.

To compare disk properties between FUors and regular YSOs, we take the Class\,II comparison sample from \citet{andrews10}, who observed 16 T Tauri disks with the SMA in the Ophiuchus star-forming region. We chose this sample because they were fitted with the same disk profile and RT models as our sample of FUors (\autoref{sec:rtmodeling}).

For a Class\,I comparison sample, we took the results of \citet{sheehan2014,sheehan2017,sheehan2017b}, who observed objects mostly in Taurus and one source in Ophiuchus. Their disk modeling employed a pure power-law surface density profile without an exponential taper. We numerically compared their surface density profiles with ours (Eqn.~\ref{eqn:sigma}), and found that using the outer disk radius $R_{\rm disk}$ from \citet{sheehan2017} as our characteristic radius $R_c$, we would get very similar surface density profiles and identical total disk masses. Therefore, in lack of any work for Class\,I disks using exactly the same model as here, the parameters from \citet{sheehan2014,sheehan2017,sheehan2017b} are the most comparable to ours.

To extend the FUor sample, we searched the literature for other FUors modeled in a way similar to ours. We found that PP~13S* from \citet{perez2010}, V883~Ori, HBC~494, V1647~Ori, and V2775~Ori from \citet{cieza2018}, and FU~Ori from \citet{perez2020} can be added to our sample, although for PP~13S* we first had to calculate $R_c$, because \citet{perez2010} normalized the surface brightness profile in a different way (their $R_t=13$\,au corresponds to $R_c=30$\,au). In the following, we work with these 17 FUors, 16 Class\,II objects, and 11 Class\,I objects.

\begin{figure}
\centering
\includegraphics[height=\columnwidth,angle=90]{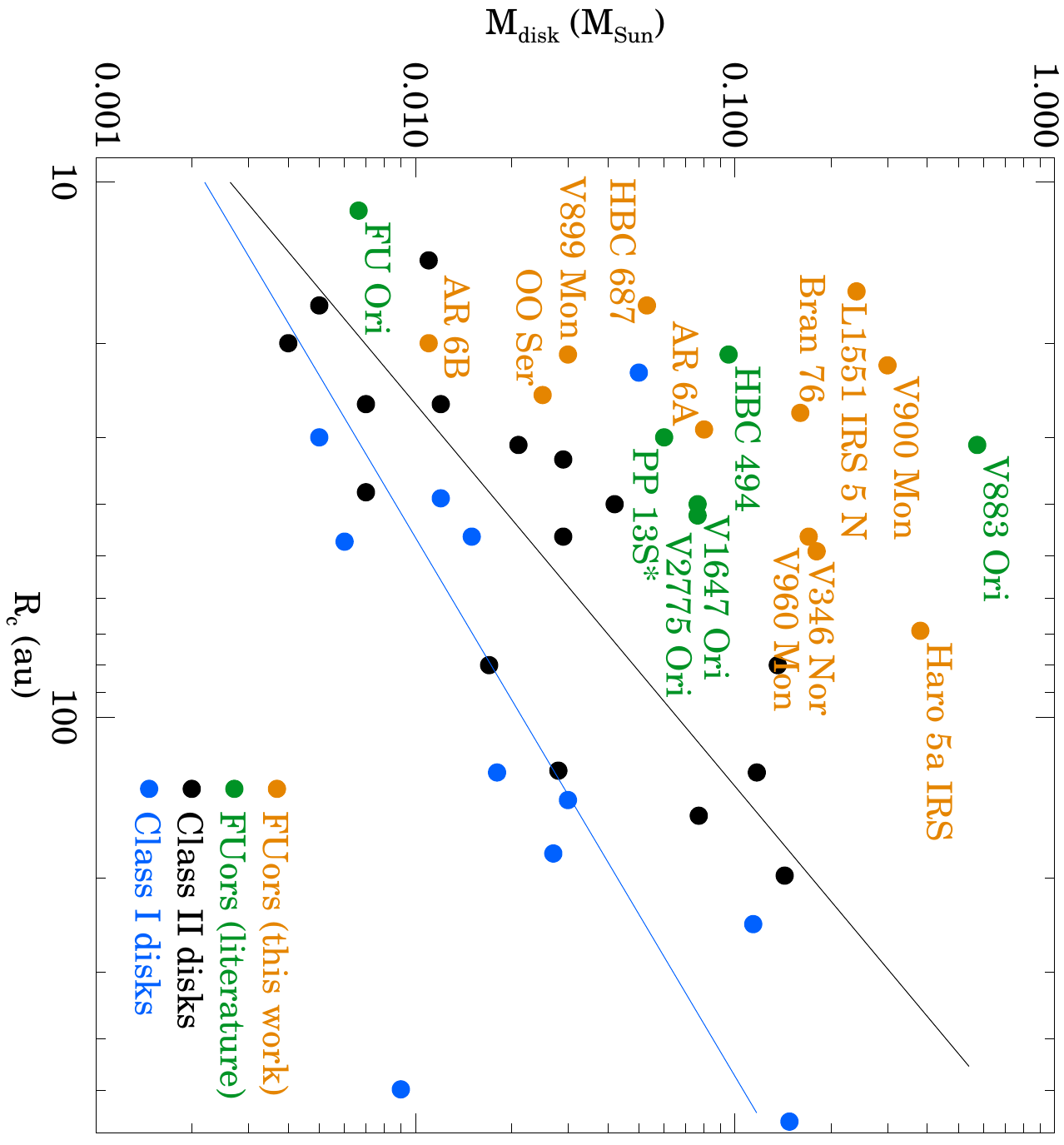}
\caption{Total disk mass (in M$_{\odot}$) as a function of characteristic radii. The orange points are the FUors presented here. The green points are FUors from  \citet{perez2010}, \citet{cieza2018}, and \citet{perez2020}. The black points are T Tauri disks from \citet{andrews2010}. The blue points are Class\,I disks from \citet{sheehan2014,sheehan2017,sheehan2017b}. Linear fits are $M_{\rm disk} \propto R_{\rm c}^{1.40}$ for Class\,II and $M_{\rm disk} \propto R_{c}^{0.99}$ for Class\,I.\label{fig:mass_radius}}
\end{figure}

\begin{figure}
\centering
\includegraphics[height=0.49\columnwidth,angle=90]{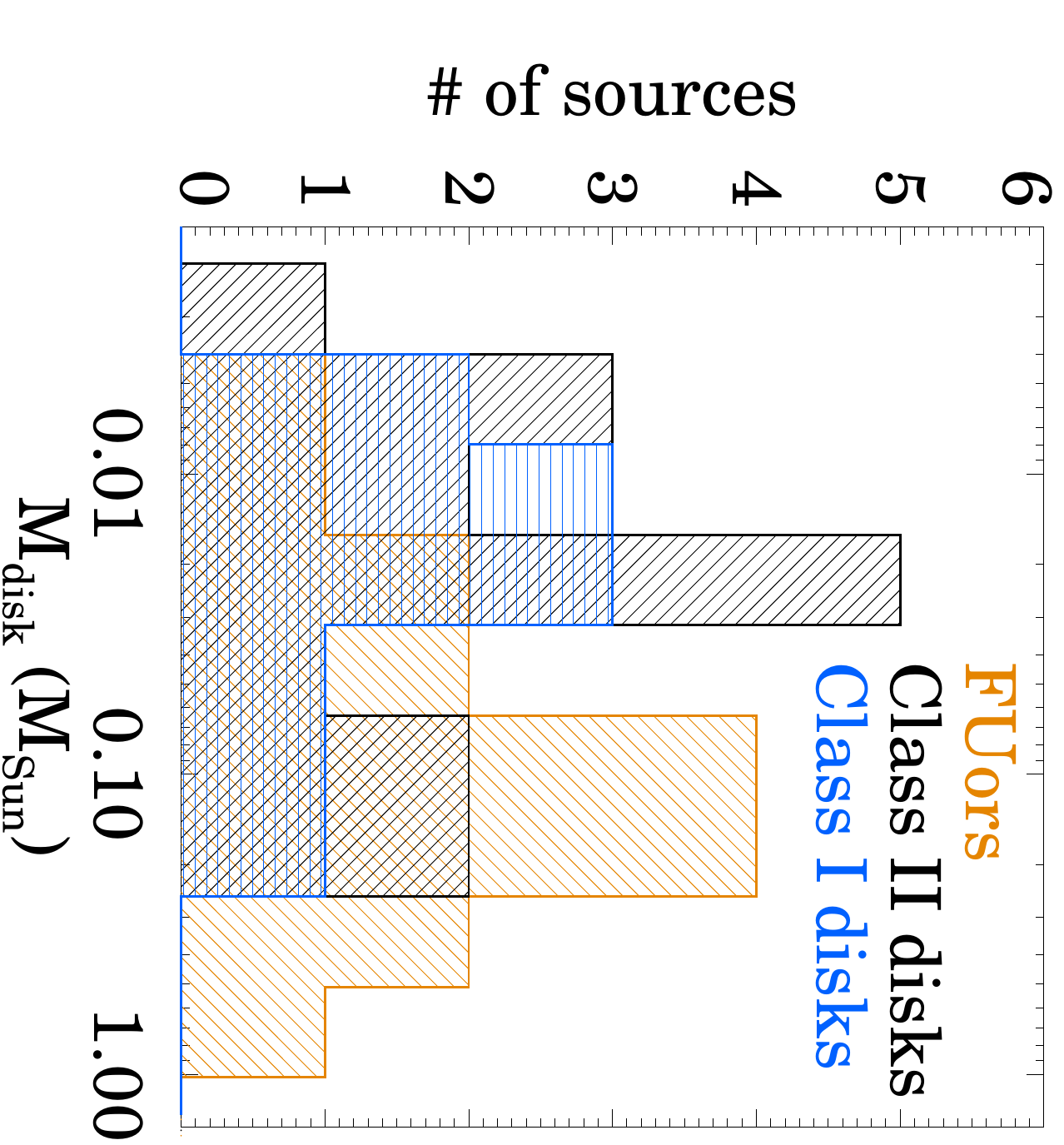}
\includegraphics[height=0.49\columnwidth,angle=90]{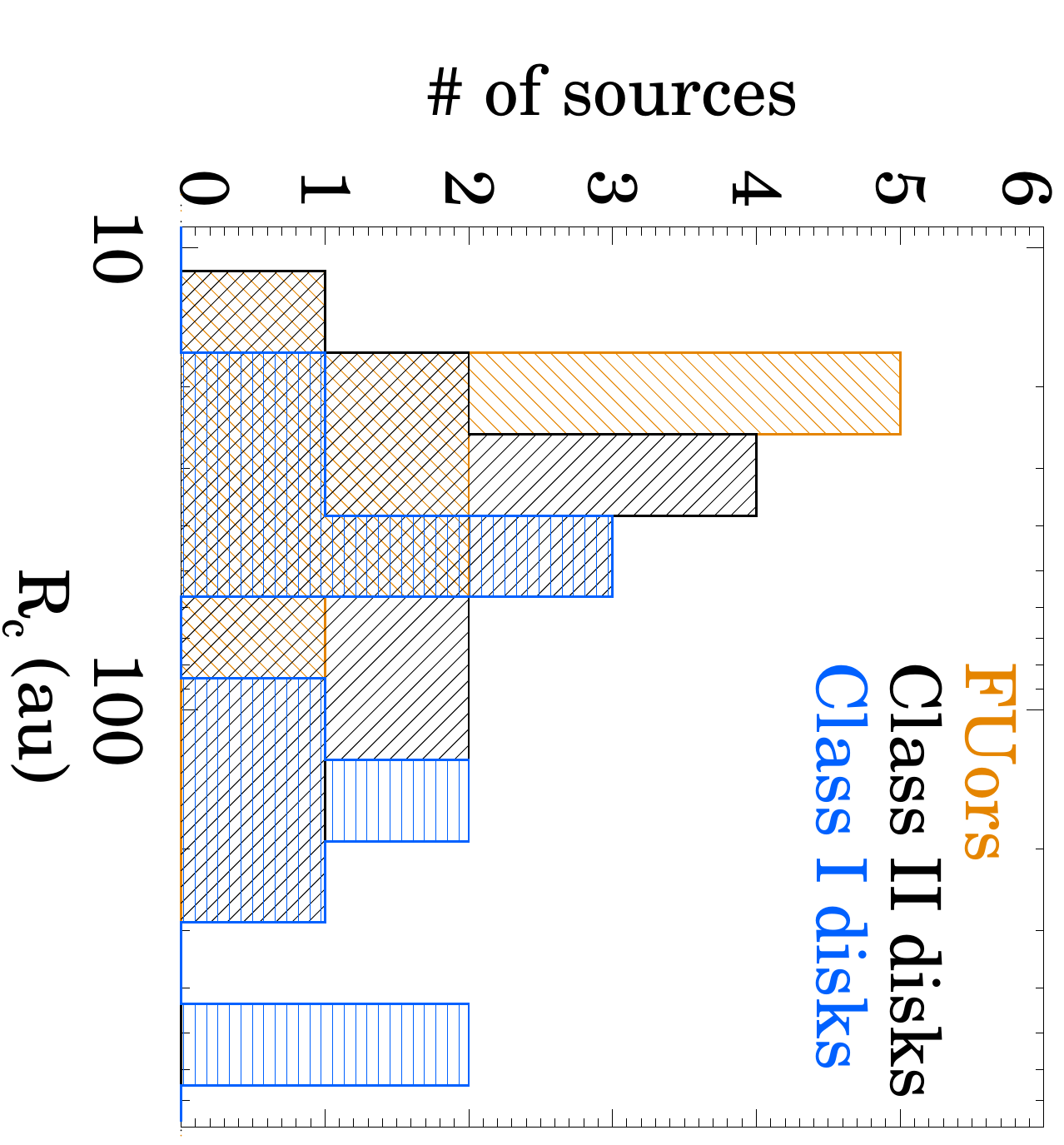}
\includegraphics[height=0.49\columnwidth,angle=90]{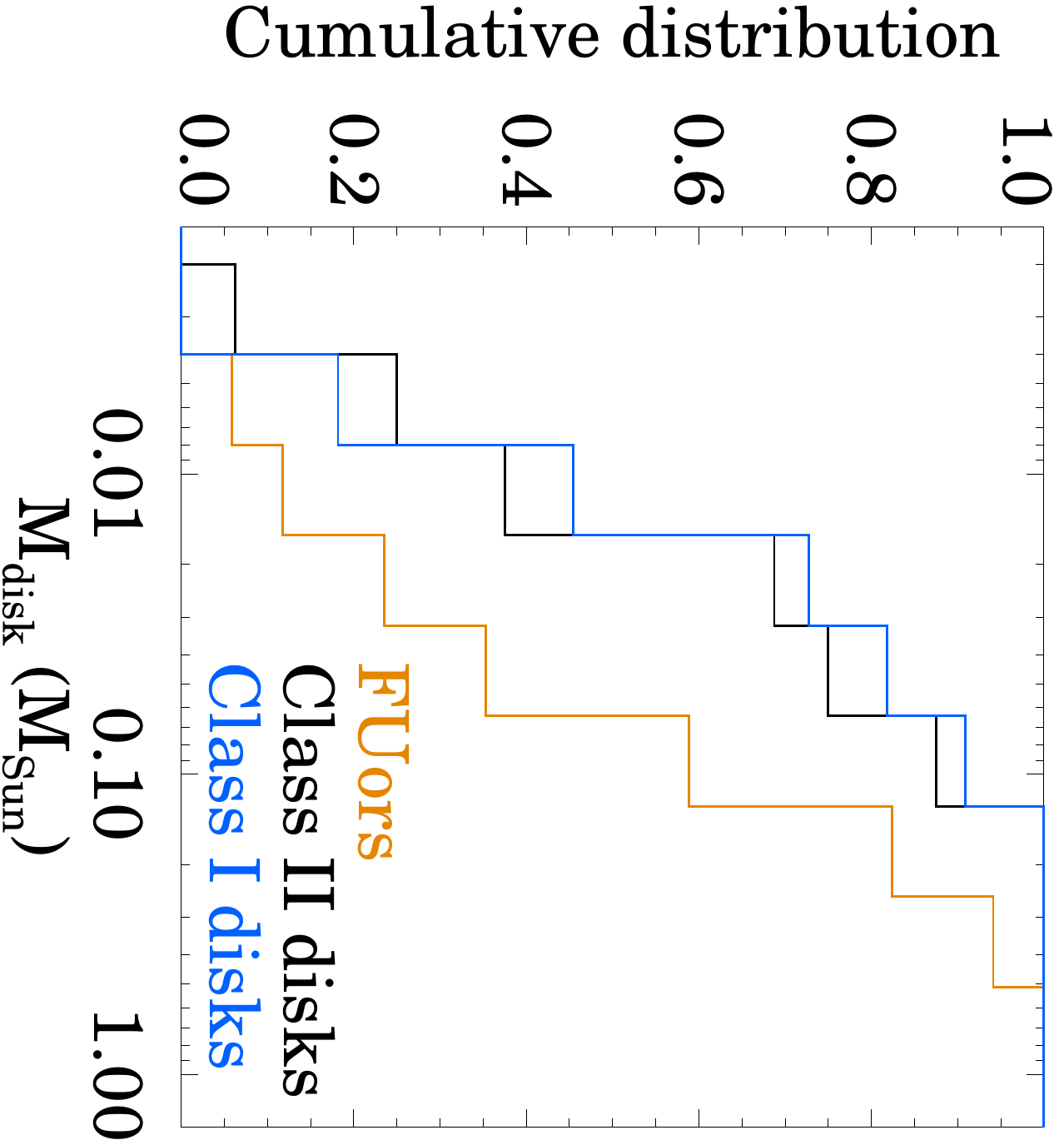}
\includegraphics[height=0.49\columnwidth,angle=90]{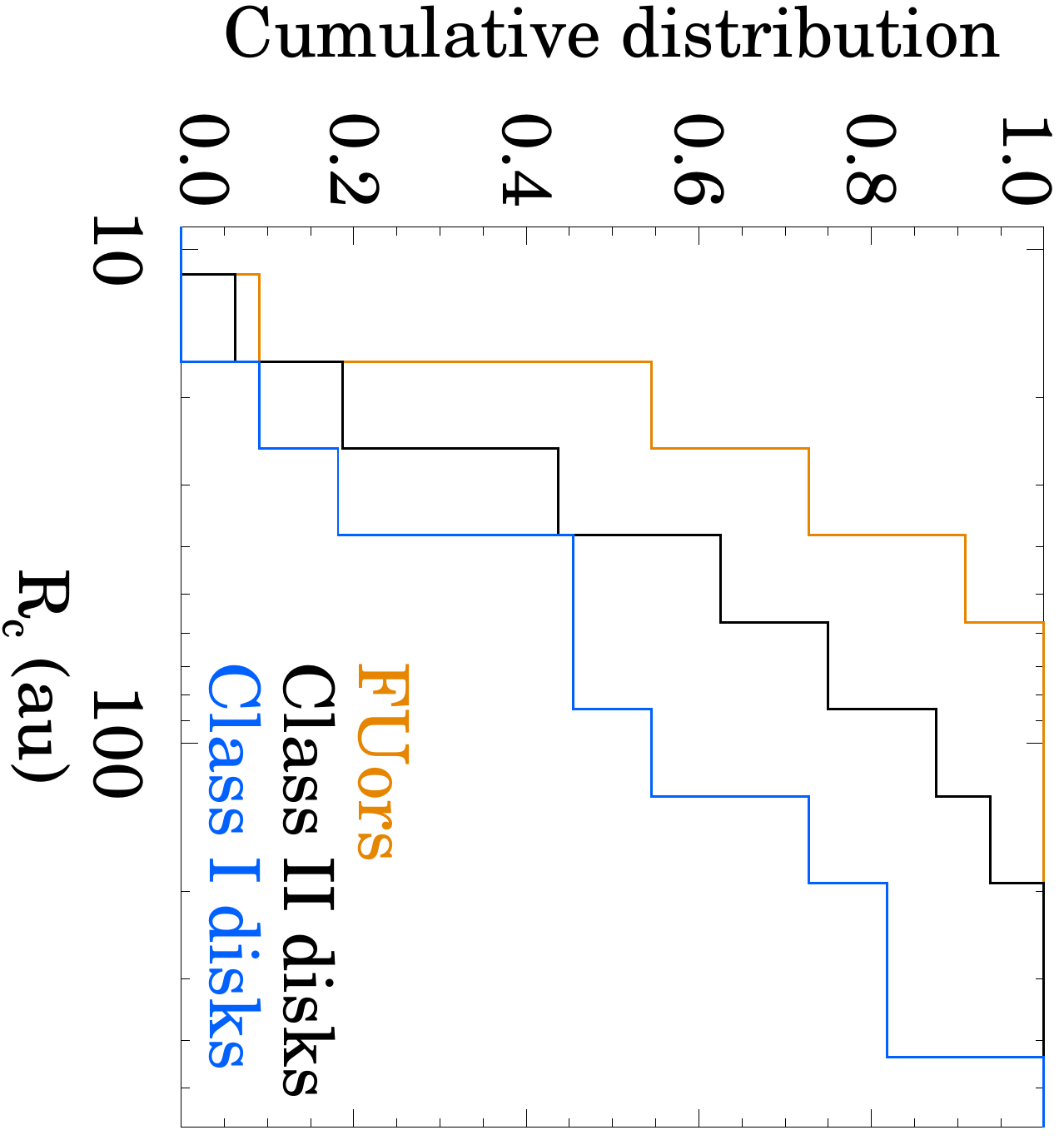}
\caption{Histograms and cumulative distributions of disk masses and disk radii from radiative transfer modeling. FUors include the sample presented here and FUors from \citet{perez2010}, \citet{cieza2018}, and \citet{perez2020}. Class\,II and Class\,I disks are the same as in \autoref{fig:mass_radius}.\label{fig:stat}}
\end{figure}

\begin{table*}
\footnotesize{
\begin{center}
\caption{P-values of various two sample statistical tests\label{tab:stat}}
\begin{tabular}{lcccccc}
\tableline\tableline
Statistical test                   & FUors/Class II & FUors/Class I & Class II/Class I & FUors/Class II & FUors/Class I & Class II/Class I \\
                                   & disk masses        & disk masses       & disk masses          & disk  radii        & disk radii        & disk radii \\
\tableline
Kolmogorov-Smirnov test            & 0.0254         & 0.0214        & 0.9997        & 0.1967         & 0.0348        & 0.3738 \\
Wilcoxon rank sum test             & 0.0069         & 0.0090        & 0.9410        & 0.0660         & 0.0016        & 0.0641 \\
Welch Two Sample t-test            & 0.0136         & 0.0123        & 0.8594        & 0.0252         & 0.0291        & 0.0872 \\
\tableline
\end{tabular}
\end{center}
}
\end{table*}

In \autoref{fig:mass_radius}, we show the total disk mass as function of the characteristic radius. The FUors presented here, along with FUors from the literature, show a clear separation both from typical T Tauri disks and from typical Class\,I disks in that they are more massive and more compact. For normal Class\,I/II disks, there is a significant correlation between disk mass and disk radius (Pearson, Spearman, and Kendall correlation statistics give coefficients in the range of 0.42--0.82, with very low probabilities of 0.0005-0.05 of the null hypothesis of no correlation). To guide the eye, we fitted a line using an Ordinary Least Squares-Bisector Regression method for these comparison samples. Class\,II disks follow a mass radius relation of $M_{\rm disk} \propto R_{\rm c}^{1.40}$ and Class\,I disks $M_{\rm disk} \propto R_{c}^{0.99}$. All FUors in \autoref{fig:mass_radius} are located above these average trends for the comparison samples. In \autoref{fig:stat}, we plotted histograms and cumulative distributions for the masses and radii of the three samples. These graphs also suggest that FUor disks are generally more massive and smaller in size than the disks of either Class\,II or Class\,I objects.

To quantify whether the mass or radius of the different samples follow the same distribution (null hypothesis), we run various two-sample tests using the R software \citep{r}. \autoref{tab:stat} shows the resulting probabilities of the null hypothesis for the Two-sample Kolmogorov-Smirnov test, the Wilcoxon rank sum exact test, and the Welch Two Sample t-test. The results suggest that the mass distribution of FUor disks are significantly different from both that of the Class\,II or Class\,I disks. The mass distribution of Class\,II and Class\,I disks, on the other hand, seem to be similar. This is supported by their cumulative distributions in the lower left panel of \autoref{fig:stat} as well. Concerning the disk radii, all three samples are different, with the FUor disks being the most compact and the Class\,I disk being the largest (see also the lower right panel of \autoref{fig:stat}).

Our results suggest that FUor disks differ from Class\,I or Class\,II disks in the distribution of their masses and radii. For the merged FUor sample used in Figures \ref{fig:mass_radius}--\ref{fig:stat}, the median FUor disk radius is 27\,au, while the median FUor disk mass is 0.08\,$M_{\odot}$. The respective values for the comparison samples are: median disk radius for Class\,I disks is 127\,au, for Class\,II disks is 40\,au, median disk mass for Class\,I is 0.02\,$M_{\odot}$ and for Class\,II disks is 0.03\,$M_{\odot}$. Therefore, while individual FUor disks may be large and not very massive, the whole sample on average is more compact (by a factor of 1.5--4.7) and more massive (by a factor of 2.9--4.4) than the disks around Class\,I or Class\,II objects.

\begin{figure}
\centering
\includegraphics[height=\columnwidth,angle=90]{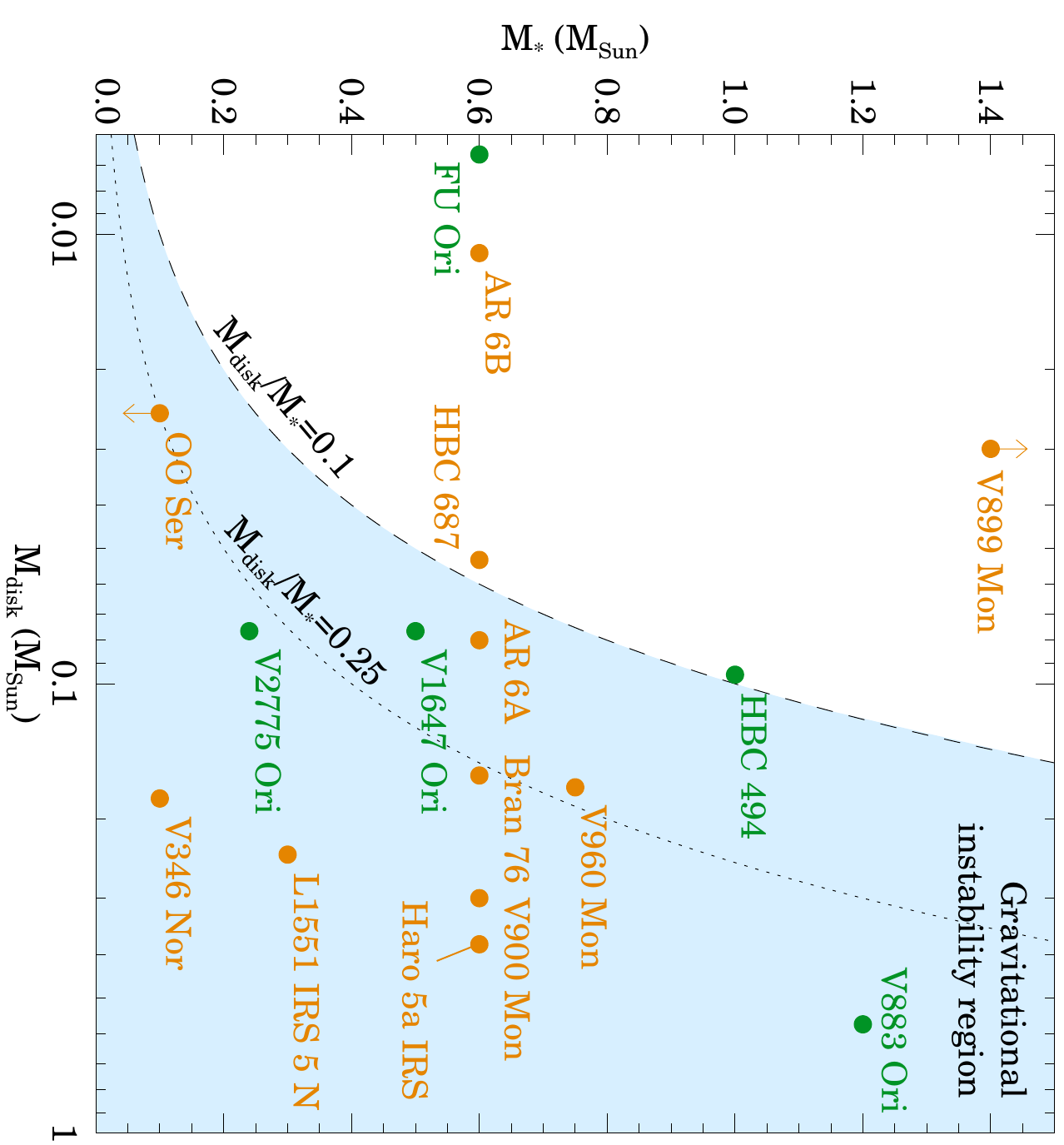}
\caption{Stellar mass as a function of disk mass (both in M$_{\odot}$). Orange dots indicate FUors from this work, green dots  are objects from the literature. Light blue shading indicates the area where the disk is relatively massive compared to the host star ($M_{\rm disk}/M_* \geq 0.1$), and thus most likely to be gravitationally unstable. The dotted curve marks the disk to steller mass ratio above which gravitational instability-induced substructures are readily detectable. V899~Mon is actually outside of the plotted region at $M_{*}=2.0\,M_{\odot}$.\label{fig:GI_plot}}
\end{figure}

\subsection{Gravitational instability in FUor disks}

\begin{deluxetable*}{cc@{}c@{}cccccc}
\tablecaption{Gravitational instability as calculated by four different metrics.\label{tab:gi}}
\tablewidth{0pt}
\tablehead{
\colhead{Name} & \colhead{$M_{\rm disk}$} & \colhead{$M_{*}$} & \colhead{$M_{\rm disk}/M_{*}>0.1$} & \colhead{$M_{Disk}/M_{*}>0.25$} & \colhead{$M_{\rm disk}/M_{*}>H/r$} & \colhead{$Q(r)<1.4$} & \colhead{$Q_{\rm min}$} & \colhead{$r(Q_{\rm min})$} \\
& \colhead{($M_{\odot}$)} & \colhead{($M_{\odot}$)} & & & & & & \colhead{(au)}}
\startdata
L1551 IRS 5 N  & 0.24  & 0.25  & Y & Y & Y & Y & 0.71 & 19 \\  
Haro 5a IRS  & 0.38  & 0.6  & Y & Y & Y & Y & 0.75 & 95 \\  
V899 Mon  & 0.030  & 2.0  & N & N & N & N & 7.4 & 17 \\  
AR 6A  & 0.080  & 0.6  & Y & N & Y & N & 2.5 & 24 \\  
AR 6B  & 0.011  & 0.6  & N & N & N & N &27 & 22 \\  
V900 Mon  & 0.30  & 0.6  & Y & Y & N & Y & 0.74 & 29 \\  
V960 Mon  & 0.17  & 0.75  & Y & N & N & Y & 0.90 & 46 \\  
Bran 76  & 0.16 & 0.6  & Y & Y & N & Y & 1.3 & 27 \\  
V346 Nor  & 0.18  & 0.1  & Y & Y & Y & Y & 0.52 & 55  \\  
OO Ser  & 0.025 & $<0.1$  & Y & Y & N & N & $<5.4$ & 29 \\  
HBC 687  & 0.053  & 0.6  & N & N & N & N & 5.6 & 17 \\  
\hline
\multicolumn{3}{c}{Total \# of gravitationally unstable disks} & 8 & 6 & 3 & 6 &
\enddata
\end{deluxetable*}

The more massive and moderately more compact nature of the FUor disks could yield insight into the source of the episodic accretion that drives the outbursts. While the exact mechanisms that cause enhanced accretion of disk material onto the protostar are still unknown, it may be triggered if the disk becomes gravitationally unstable, possibly in combination with magnetorotational instability \citep{armitage2001, lodato2004, vorobyov2006, vorobyov2010, boley2008, zhu2009, martin2011, bae2014, kadam2019}. This scenario most likely occurs in relatively massive disks when $M_{\rm disk}/M_{*}{\geq}0.1$ \citep{kratter2016}. The actual criterion is closer to $M_{\rm disk}/M_{*}{\geq}H/r$, i.e., if the $M_{\rm disk}/M_{*}$ ratio is greater than or equal to the disk aspect ratio (the disk scale height over the characteristic radius). More accurately,
gravitational instability occurs if the value of Toomre's $Q$ parameter is $Q = c_s \Omega/\pi G \Sigma \lesssim 1.4$, where $c_s$ is the sound speed, $\Omega$ is the epicyclic frequency for a Keplerian disk, and $\Sigma$ is the column density at a given radius in the disk. Gravitationally unstable disks are expected to develop substructures like spiral arms \citep[e.g.,][]{durisen2007,kratter2016}. Numerical simulations by \citet{hall2019} indicate that gravitational instability-induced spiral structures are only detectable when $M_{\rm disk}/M_{*}{\geq}0.25$.

To check these criteria for our sample, we collected stellar mass estimates for FUors from the literature. We used $<0.1\,M_{\odot}$ for OO Ser \citep{hodapp1996}, $0.3\,M_{\odot}$ for L1551~IRS~5 \citep{osorio2003}, $0.75\,M_{\odot}$ for V960~Mon \citep{kospal2015}, $2\,M_{\odot}$ for V899\,Mon \citep{ninan2015}, $0.1\,M_{\odot}$ for V346~Nor \citep{kospal2017c}, $0.6\,M_{\odot}$ for FU~Ori \citep{perez2020}. For objects without published stellar mass, we assumed the same value as for FU~Ori, 0.6$\,M_{\odot}$. For all calculations, the RT disk masses were used. To calculate c$_{s}$ and $\Omega$, we used the RT temperature and density profiles. \autoref{tab:gi} shows whether a disk satisfies any of the four criteria mentioned above and the number of disks that are gravitationally unstable based on these metrics.

We plotted the stellar masses as a function of the disk mass in \autoref{fig:GI_plot}. Here, we complemented our sample with objects from the literature. The $M_{\rm disk}/M_{*}{\geq}0.1$ and $M_{\rm disk}/M_{*}{\geq}0.25$ criteria (for likely instability and for the appearance of substructures, respectively) are indicated by dashed and dotted curves in this figure. Interestingly, while some FUors clearly satisfy the stricter criterion, we do not detect such structures in our targets. The reason for this may be that the spatial resolution of our ALMA observations is insufficient to resolve the expected substructures in these relatively distant objects. Alternatively, the ongoing burst might have sent out an expansion wave through the disk and smeared out the preexisting spiral structure, as recently proposed by \citet{vorobyov2020}.

\autoref{fig:GI_plot} shows that 11 out of a sample of 17 FUors (65\%) are in the gravitational instability region. Even if we only consider objects with published stellar masses, there are 7 FUors out of 10 (70\%) that may have gravitationally unstable disks, suggesting that gravitational instability may be an important factor in driving the outbursts. On the other hand, V899~Mon, FU~Ori, and possibly AR~6A and HBC~687 as well, seem to harbor gravitationally stable disks, and triggering their outbursts may need some other mechanisms.

\subsection{Implications for the ubiquity of the FUor phenomenon}

We discussed previously that FUor disks are on average more compact and more massive than the disks around Class\,I or Class\,II objects. This conclusion has important implications and raises interesting questions. One possible interpretation of these results is that  there is a phase of the circumstellar disk evolution when the disk structure temporarily deviates from the typical one, forms a new structure in which it can produce outbursts, then relaxes back to a normal disk structure.

Alternatively, FUor eruptions may happen in normal disks, but the outburst itself (the increased accretion heating, the enhanced wind activity, etc.) transforms the disk structure from normal into the special one we observe around eruptive stars. In these cases, it is a question what physical mechanism can make the disks smaller and their masses larger (especially in later evolutionary phases when replenishment from the envelope is not significant any more).

Finally, it may be possible that some disks are born with smaller radius and larger mass than usual, and later they follow a particular evolutionary path that leads to episodic eruptions. Numerical simulations suggest that this may be the consequence of delayed infall from a slowly rotating envelope \citep{vorobyov2015b}. If FUor disks are born special and follow a special evolutionary path, we would need to accept that not all YSOs go through FUor eruptions.

While it is often assumed that all YSOs undergo FUor eruptions, our results raise the possibility that they form instead a special class of young stellar objects defined by their unusual disk structure and enhanced disk accretion. If so, then members of this special class are relatively rare among young stars. \citet{contreraspena2019} analyzed a large sample of 15\,000 YSOs and identified only 9 FUors, suggesting that at any given time, only 0.06\% of young stars are in FUor outburst. A somewhat higher rate was reported for the population of YSOs in the Orion Molecular Clouds by \citet{fischer2019}, who found two outbursting sources out of 319 protostars (0.6\%). The true incidence rate of FUors can be much higher if we take into account their duty cycle, i.e., the fraction of the life of a FUor-type object spent in outburst.

One way to determine the true fraction of eruptive stars among YSOs is to monitor them and discover new outbursts. All-sky monitoring surveys like Gaia and WISE gave a new impetus to this field of study \citep[e.g.,][]{szegedi-elek2020,hillenbrand2018,hillenbrand2019,hillenbrand2021}. Another way would be to explore the accretion properties and structure of disks falling above the Class\,I/II relations in \autoref{fig:mass_radius}.

Less than one percent is a rather low rate for actual outbursting FUors, but these objects may still be important players in the star formation process. Considering that their accretion rate is $\sim$1000 times higher than of normal YSOs ($10^{-4}$ instead of $10^{-7}$, \citealt{hk96}), the  matter accreted in the young eruptive stars may be comparable or even higher to the total mass accreted in the whole known YSO population.


\section{Summary and Conclusions}
\label{sec:summary}

We presented an ALMA millimeter continuum study of FUors, a class of low-mass pre-main sequence stars undergoing accretion-related outbursts. Our sample of 10 objects contains a large part of all FUors visible from the ALMA site. We mapped the thermal dust continuum emission at 1.33\,mm using the 7\,m array and two different configurations of the 12\,m array. This way we recover emission at the widest possible range of spatial scales from the typical beam size of 0$\farcs$15 to the largest recoverable scale of 5$\farcs$9. We detected all of the targeted FUors and several neighboring sources as well. In most cases the detected emission is bright, centrally peaked, and rather compact, and we attribute them to the circumstellar disks. To obtain basic geometrical parameters and the integrated millimeter fluxes of the sources, we first fitted 2D Gaussians to the images. As a next step, we fitted an analytical disk model of a power law brightness distribution with an exponential taper in the visibility space using an Markov chain Monte Carlo method. Finally, we computed radiative transfer (RT) disk models with RADMC-3D, and fitted them in the image space again with Markov chain Monte Carlo method.

The geometrical parameters from the different models are generally in good agreement within the uncertainties. However, the RT modeling gives disk masses significantly larger than what is obtained from the measured millimeter fluxes assuming  optically thin emission, suggesting that the FUor disks are optically thick at 1.33\,mm. From the RT modeling, we obtained disk masses  between 0.011 and 0.38$\,M_{\odot}$ and disk characteristic radii between 16 and 69\,au. We complemented our sample with FUors from the literature and compared their masses and radii with comparison samples of normal, non-eruptive  Class\,I and Class\,II objects modeled in a similar way with RT codes. The comparison revealed that FUor disks are typically a factor of 2.9--4.4 more massive and a factor of 1.5--4.7 smaller in size than the disks of Class\,I/II objects. Using the criterion of $M_{\rm disk}/M_{*}{\geq}0.1$, about 65--70\% of the FUor disks may be gravitationally unstable.

It remains unclear whether all normal protostellar disks could develop properties that we observe for the FUor sample discussed here, which would imply that episodic accretion during the pre-main sequence stellar evolution may be common. Our main finding is that FUor disks are generally massive and compact may imply that they form a special class or evolutionary path.

There are several possible future directions that may provide better insights into the physics of young eruptive stars than what is possible at the moment. RT modeling of larger samples of YSO disks will enable more robust comparison statistics that we could do here. Deeper and higher spatial resolution millimeter interferometric observations may better trace the disk-envelope connection and might reveal substructures in the disks related to gravitational instability. Observations at multiple bands, including longer wavelengths with the planned Band 1 and 2 receivers, will help constrain dust properties and optical depth effects. Targeting the appropriate molecular tracers will allow to study possible outflow activity and interesting chemical processes related to the enhanced luminosity during the outburst. Thanks to several ongoing and planned monitoring surveys, new eruptive star candidates will be discovered, whose follow-up may hold the key to understanding episodic accretion and its effects.


\vspace*{5mm}

\begin{acknowledgments}
We thank the anonymous reviewer for useful comments that helped us to improve the paper. This project has received funding from the European Research Council (ERC) under the European Union's Horizon 2020 research and innovation programme under grant agreement No 716155 (SACCRED). This paper makes use of the following ALMA data: ADS/JAO.ALMA\#2016.1.00209.S. ALMA is a partnership of ESO (representing its member states), NSF (USA) and NINS (Japan), together with NRC (Canada) and NSC and ASIAA (Taiwan) and KASI (Republic of Korea), in cooperation with the Republic of Chile. The Joint ALMA Observatory is operated by ESO, AUI/NRAO and NAOJ. On behalf of the SACCRED project we thank for the usage of MTA Cloud (https://cloud.mta.hu/) that significantly helped us achieving the results published in this paper. M.T. is supported by the Ministry of Science and Technology (MoST) of Taiwan (grant No.~106-2119-M-001-026-MY3, 109-2112-M-001-019). H.B.L. and M.T. and are supported by MoST of Taiwan 108-2923-M-001-006-MY3 for the Taiwanese-Russian collaboration project. E.V. acknowledges support from  the Russian Fund for Fundamental Research, Russian-Taiwanese project 19-52-52011.
\end{acknowledgments}

\vspace*{5mm}

\facilities{ALMA}
\software{CASA \citep{mcmullin2007}, Galario \citep{Tazzari18}, emcee \citep{ForemanMackey13}, RADMC-3D \citep{dullemond2012radmc}, radmc3dPy}

\appendix 


\clearpage

\section{Visibility space fitting results}
\label{sec:app_a}

\begin{figure*}[!ht]
\centering
\includegraphics[width=\textwidth]{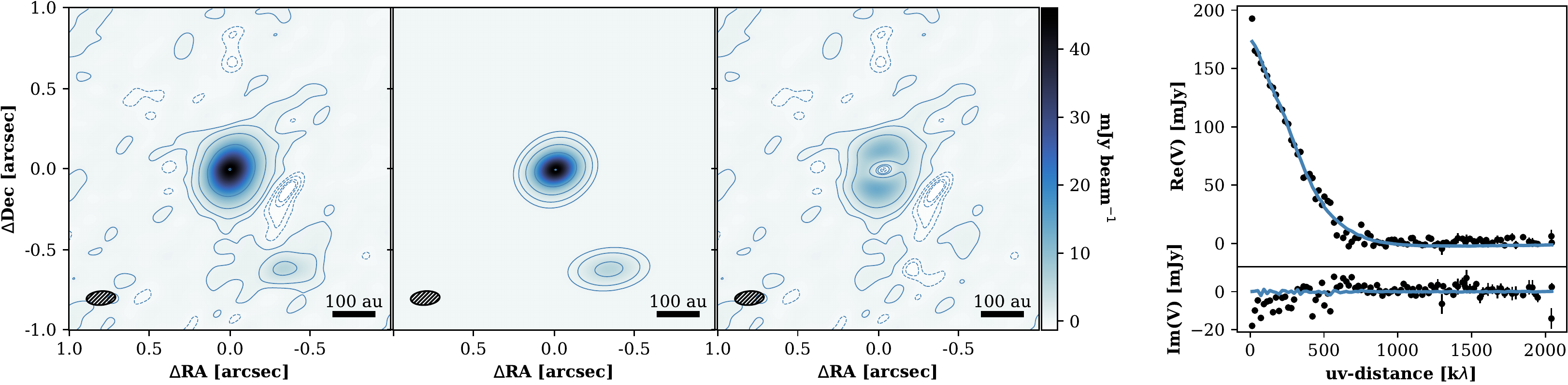}
\caption{Maps and visibility fitting of Haro~5a~IRS. Left panel is the clean map with uniform weighting, center left panel is the best fit model convolved with a 2D Gaussian representing the clean beam and the center right panel is the residual map. For the three panels in the image space the contours are at $-$10, $-$8, $-$7, $-$5, $-$3, 3, 9, 31, 103 and 334$\sigma$ with $\sigma$ = 138 $\mu$Jy. In the right panel, the black points show the observed visibility components plotted against the \textit{uv} distance, and the blue line shows the best fitted model. The \textit{uv} distances of the observed and modeled visibilities have been phase shifted to the best-fitted $\Delta$RA and $\Delta$Dec, and deprojected by $i$ and PA.\label{fig:vis_haro5airs}}
\end{figure*}

\begin{figure*}[!ht]
\centering
\includegraphics[width=\textwidth]{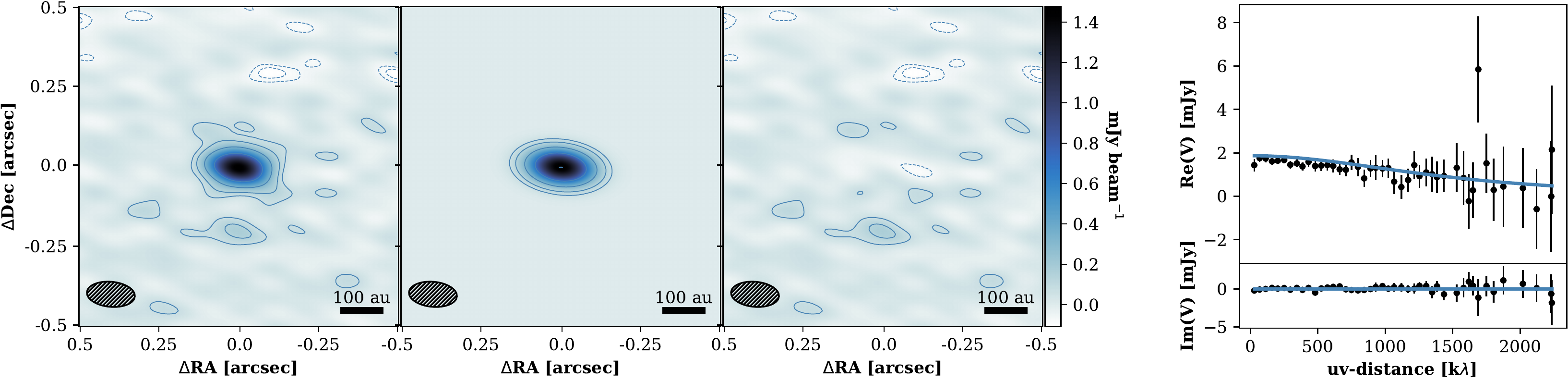}
\caption{Same as \autoref{fig:vis_haro5airs} but for V899~Mon. The contours are at $-$5, $-$4, $-$3, 3, 6, 14, 30 and 64$\sigma$ with $\sigma$ = 23 $\mu$Jy.\label{fig:vis_v899mon}}
\end{figure*}

\begin{figure*}[!ht]
\centering
\includegraphics[width=\textwidth]{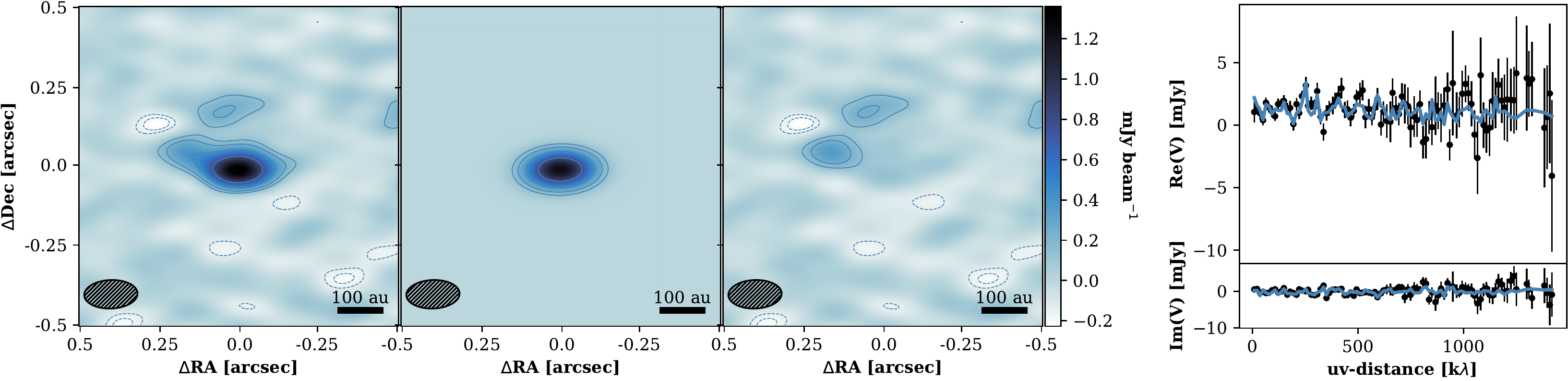}
\caption{Same as \autoref{fig:vis_haro5airs} but for AR~6A. The contours are at $-$5, $-$4, $-$3, 3, 5, 9, 17 and 30$\sigma$ with $\sigma$ = 46 $\mu$Jy.\label{fig:vis_ar6a}}
\end{figure*}

\begin{figure*}[!ht]
\centering
\includegraphics[width=\textwidth]{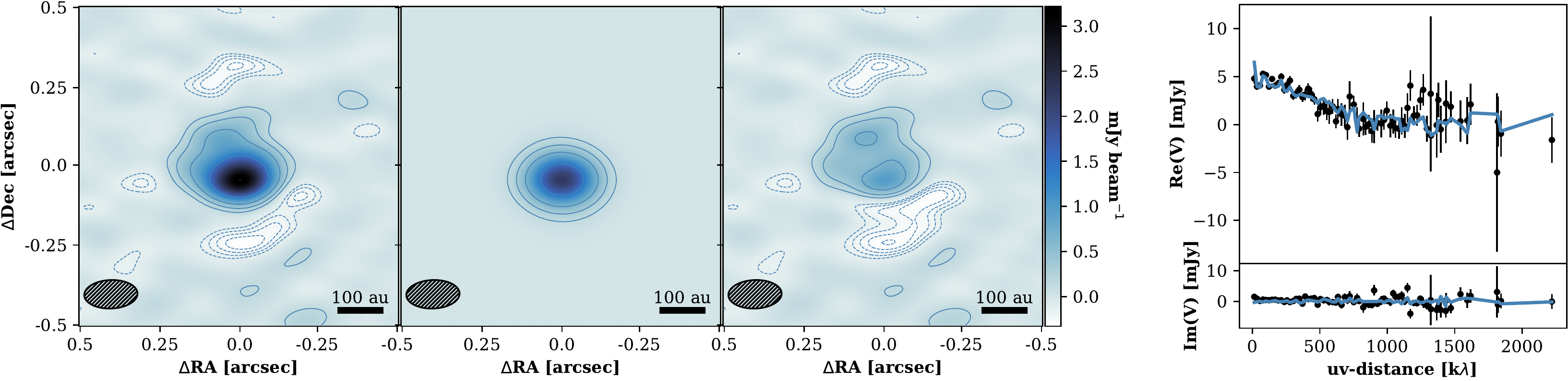}
\caption{Same as \autoref{fig:vis_haro5airs} but for AR~6B. The contours are at $-$7, $-$6, $-$5, $-$4, $-$3, 3, 7, 14, 32 and 70$\sigma$ with $\sigma$ = 46 $\mu$Jy.\label{fig:vis_ar6b}}
\end{figure*}

\begin{figure*}[!ht]
\centering
\includegraphics[width=\textwidth]{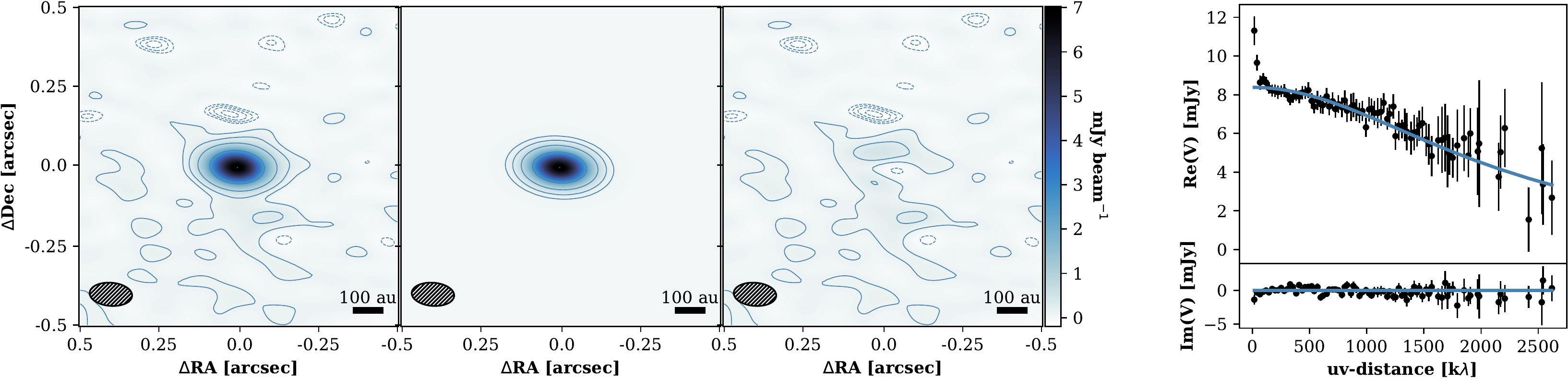}
\caption{Same as \autoref{fig:vis_haro5airs} but for V900~Mon. The contours are at $-$6, $-$5, $-$4, $-$3, 3, 9, 26, 78 and 232$\sigma$ with $\sigma$ = 30 $\mu$Jy.\label{fig:vis_v900mon}}
\end{figure*}

\begin{figure*}[!ht]
\centering
\includegraphics[width=\textwidth]{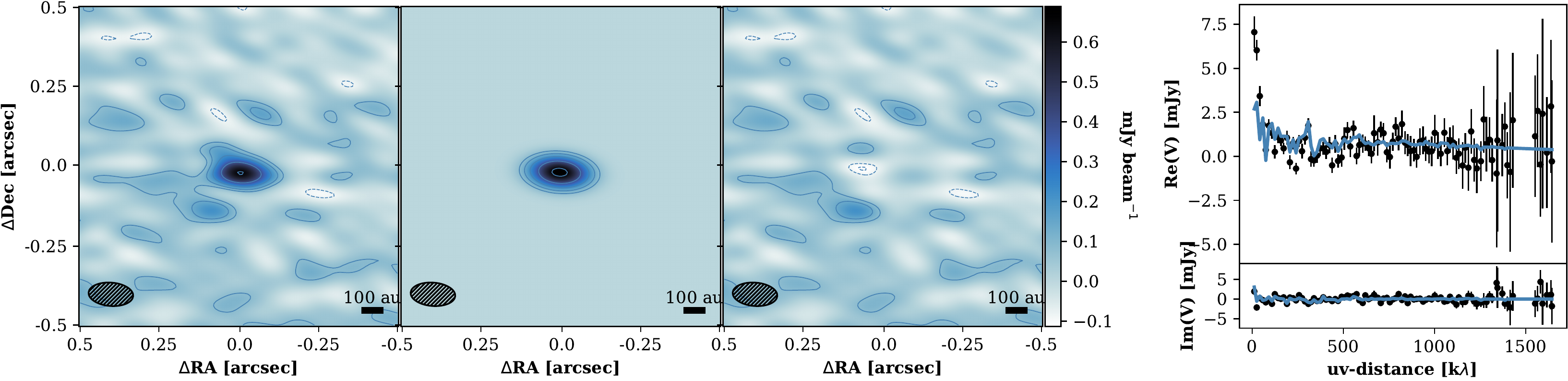}
\caption{Same as \autoref{fig:vis_haro5airs} but for V960~Mon. The contours are at $-$4, $-$3, 3, 5, 8, 14 and 23$\sigma$ with $\sigma$ = 28 $\mu$Jy.\label{fig:vis_v960mon}}
\end{figure*}

\begin{figure*}[!ht]
\centering
\includegraphics[width=\textwidth]{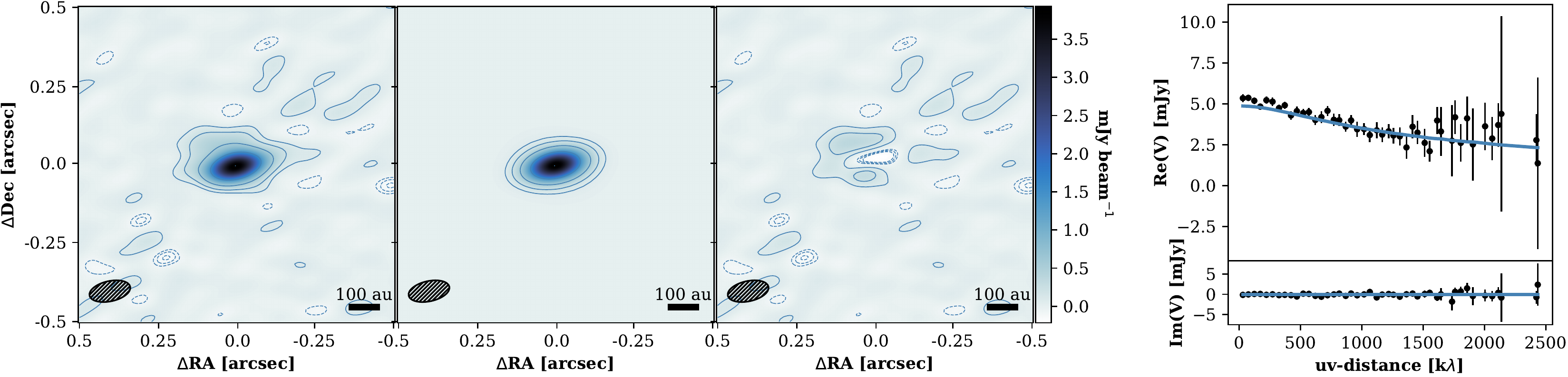}
\caption{Same as \autoref{fig:vis_haro5airs} but for Bran~76. The contours are at $-$5, $-$4, $-$3, 3, 8, 21, 55 and 146$\sigma$ with $\sigma$ = 27 $\mu$Jy.\label{fig:vis_bran76}}
\end{figure*}

\begin{figure*}[!ht]
\centering
\includegraphics[width=\textwidth]{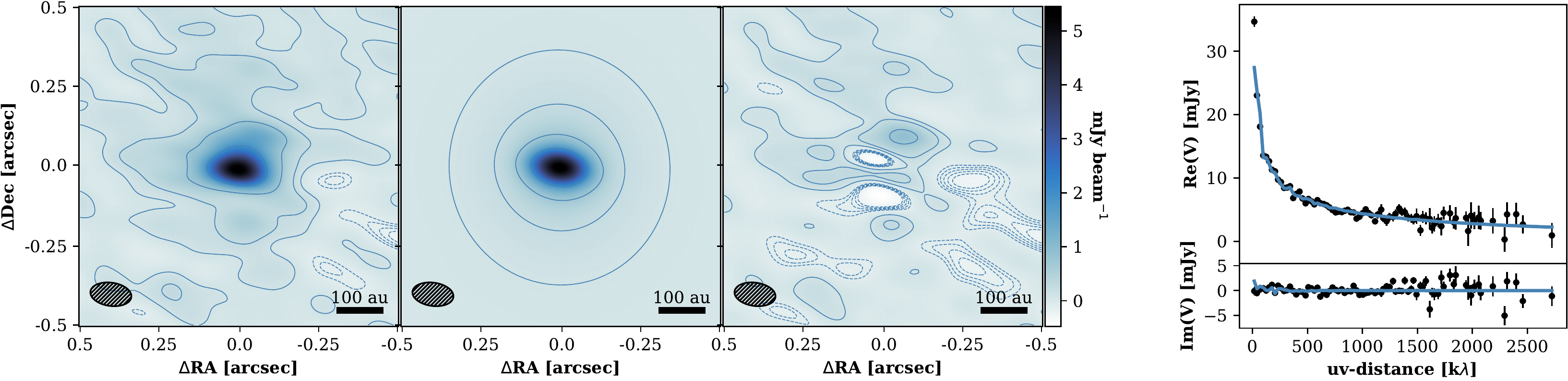}
\caption{Same as \autoref{fig:vis_haro5airs} but for V346~Nor. The contours are at $-$7, $-$6, $-$5, $-$4, $-$3, 3, 8, 23, 63 and 174$\sigma$ with $\sigma$ = 31 $\mu$Jy.\label{fig:vis_v346nor}}
\end{figure*}

\begin{figure*}[!ht]
\centering
\includegraphics[width=\textwidth]{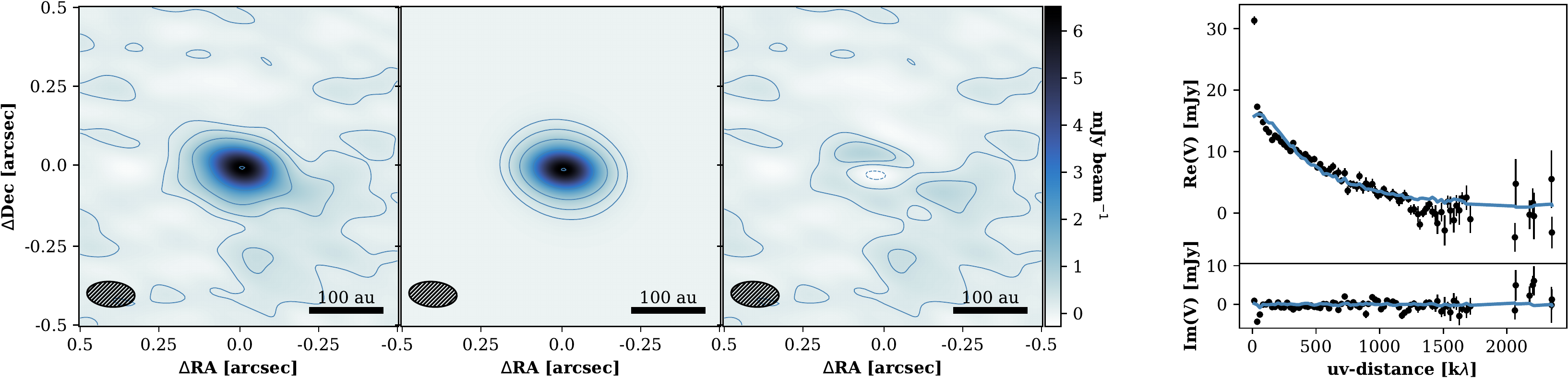}
\caption{Same as \autoref{fig:vis_haro5airs} but for OO~Ser. The contours are at $-$3, 3, 7, 17, 42 and 101$\sigma$ with $\sigma$ = 64 $\mu$Jy.\label{fig:vis_ooser}}
\end{figure*}

\begin{figure*}[!ht]
\centering
\includegraphics[width=\textwidth]{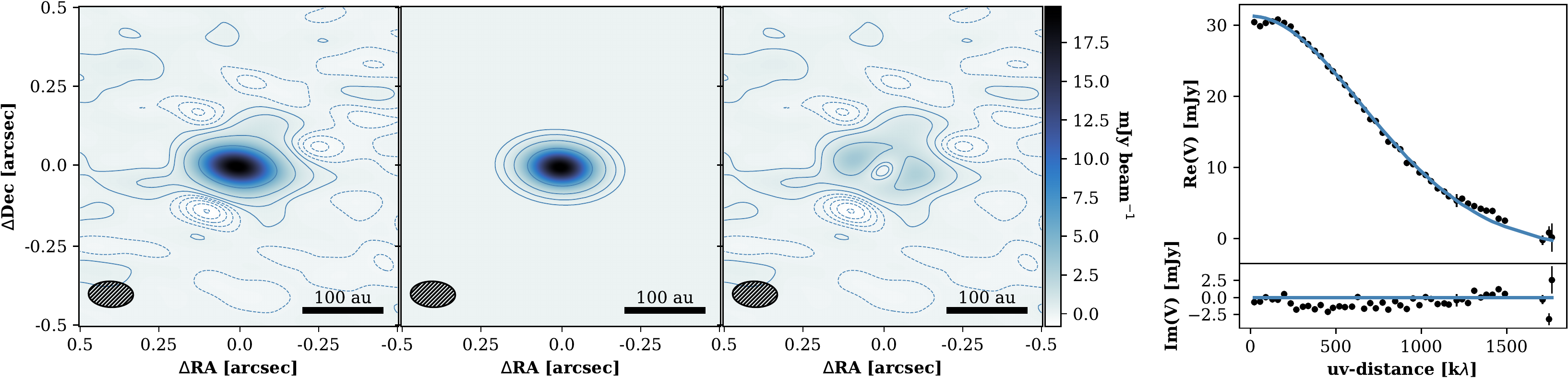}
\caption{Same as \autoref{fig:vis_haro5airs} but for HBC~687. The contours are at $-$19, $-$15, $-$11, $-$7, $-$3, 3, 11, 38, 135 and 478$\sigma$ with $\sigma$ = 42 $\mu$Jy.\label{fig:vis_parsamian21}}
\end{figure*}

Figures \ref{fig:vis_haro5airs}--\ref{fig:vis_parsamian21} show the observed clean maps, the best fit models convolved with the clean beam, the residual maps, and the observed and model visibilities as a function of $uv$ distance.

\clearpage

\section{RADMC modelling results}
\label{sec:app_b}

\begin{figure}[!ht]
\includegraphics[angle=0,width=\textwidth,trim=20 30 30 50,clip]{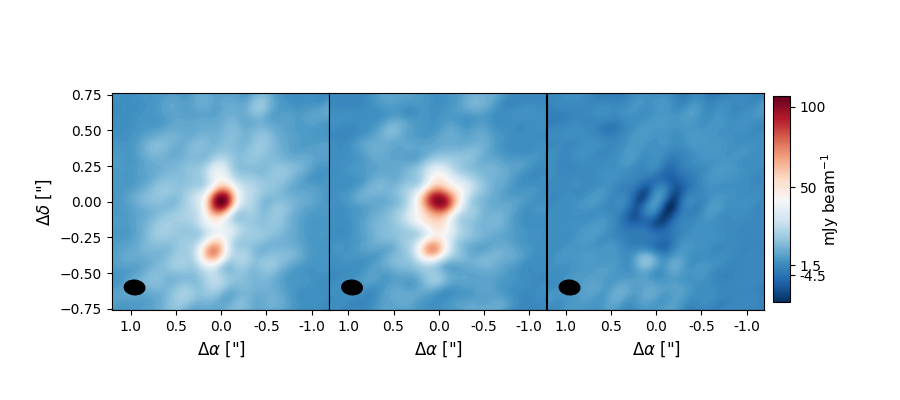}
\caption{RADMC-3D model fitting of L1551~IRS~5~N and S. Left panel is the dirty map, the central panel is the best fit model convolved with the dirty beam, and the right panel is the residual map.\label{fig:L1551}}
\end{figure}

\begin{figure}[!ht]
\includegraphics[angle=0,width=\textwidth,trim=30 30 30 50,clip]{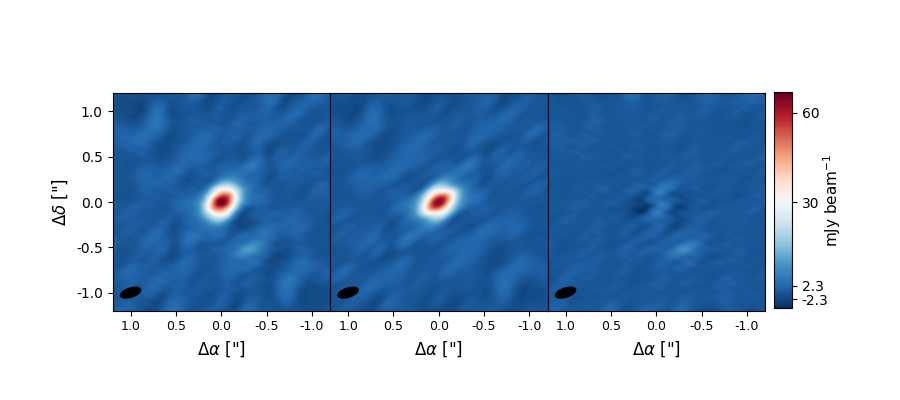}
\caption{Same as \autoref{fig:L1551} but for Haro~5A~IRS.\label{fig:Haro5A}}
\end{figure}

\begin{figure}[!ht]
\includegraphics[angle=0,width=\textwidth,trim=30 30 30 50,clip]{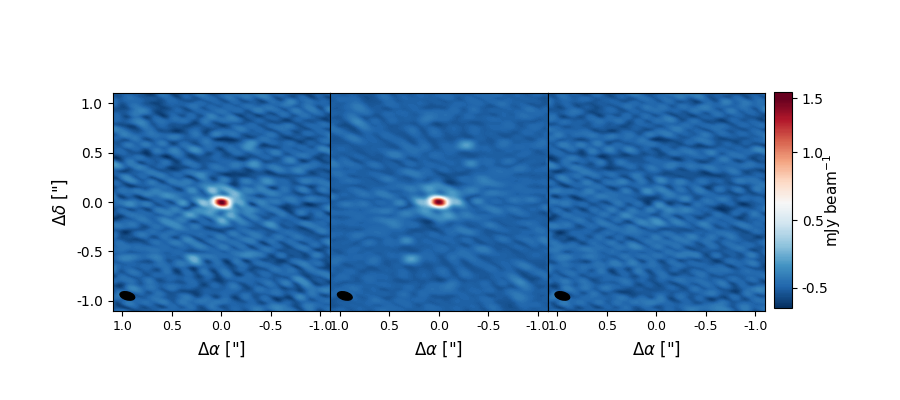}
\caption{Same as \autoref{fig:L1551} but for V899~Mon.\label{fig:V899Mon}}
\end{figure}

\begin{figure}[!ht]
\includegraphics[angle=0,width=\textwidth,trim=30 30 30 50,clip]{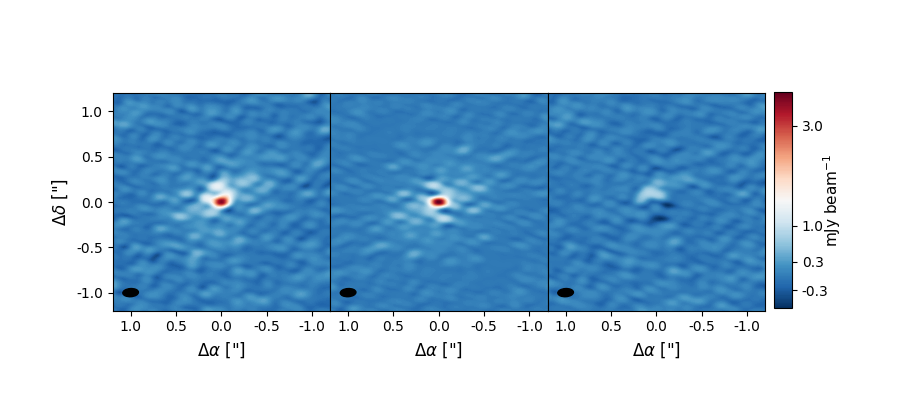}
\caption{Same as \autoref{fig:L1551} but for AR~6A.\label{fig:AR6A}}
\end{figure}

\begin{figure}[!ht]
\includegraphics[angle=0,width=\textwidth,trim=30 30 30 50,clip]{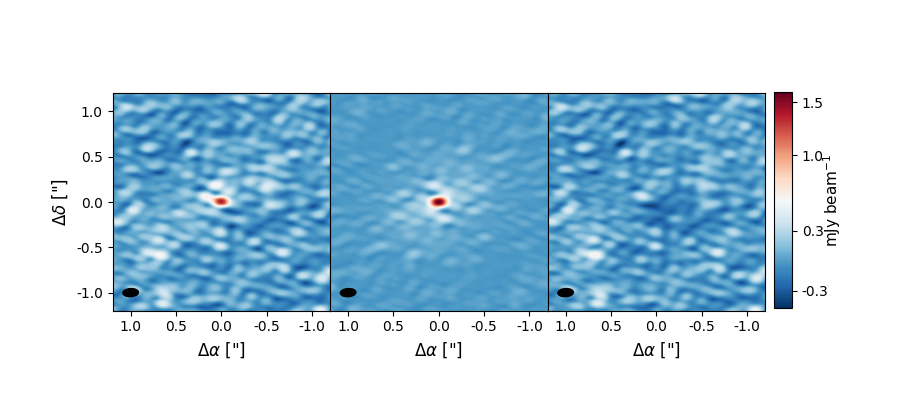}
\caption{Same as \autoref{fig:L1551} but for AR~6B.\label{fig:AR6B}}
\end{figure}

\begin{figure}[!ht]
\includegraphics[angle=0,width=\textwidth,trim=30 30 30 50,clip]{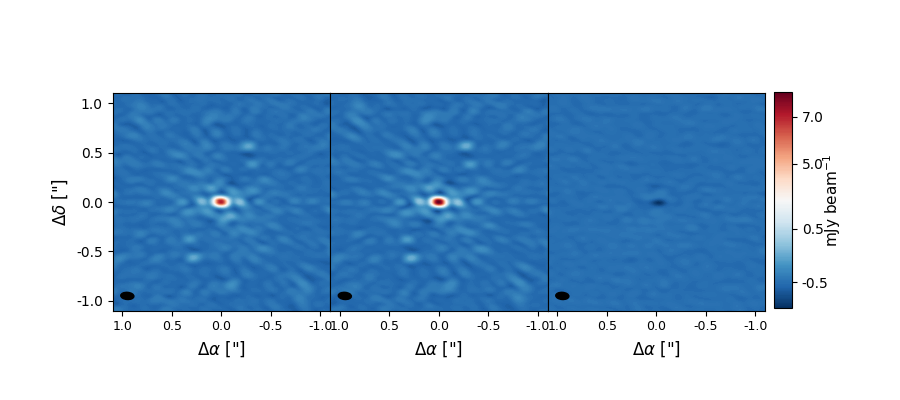}
\caption{Same as \autoref{fig:L1551} but for V900~Mon.\label{fig:V900Mon}}
\end{figure}

\begin{figure}[!ht]
\includegraphics[angle=0,width=\textwidth,trim=30 30 30 50,clip]{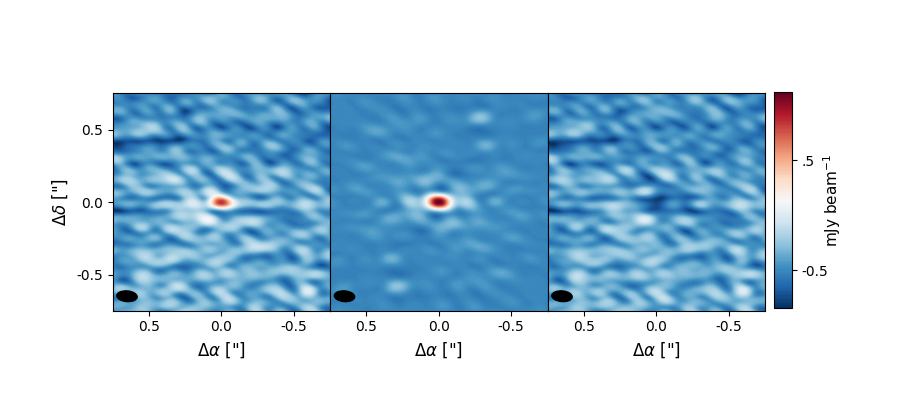}
\caption{Same as \autoref{fig:L1551} but for V960~Mon.\label{fig:V960Mon}}
\end{figure}

\begin{figure}[!ht]
\includegraphics[angle=0,width=\textwidth,trim=30 30 30 50,clip]{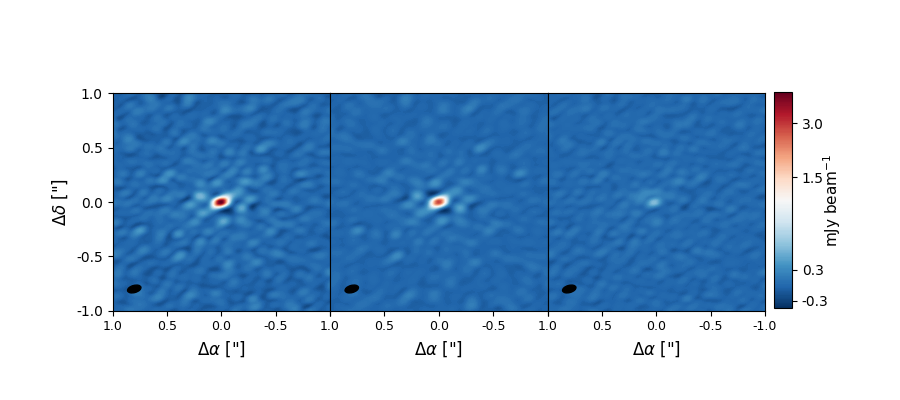}
\caption{Same as \autoref{fig:L1551} but for Bran~76.\label{fig:Bran76}}
\end{figure}

\begin{figure}[!ht]
\includegraphics[angle=0,width=\textwidth,trim=30 30 30 50,clip]{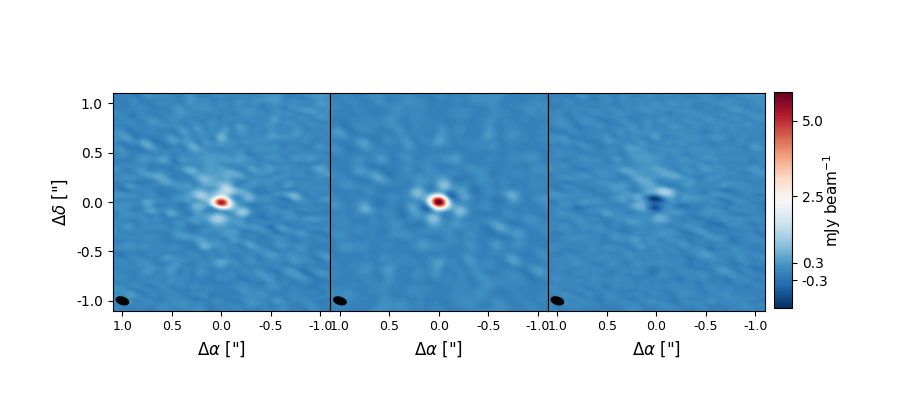}
\caption{Same as \autoref{fig:L1551} but for V346~Nor.\label{fig:V346Nor}}
\end{figure}

\begin{figure}[!ht]
\includegraphics[angle=0,width=\textwidth,trim=30 30 30 50,clip]{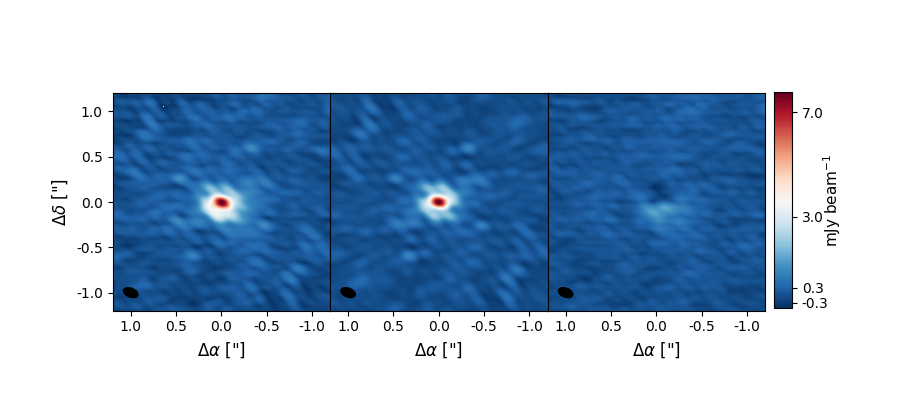}
\caption{Same as \autoref{fig:L1551} but for OO~Ser.\label{fig:OOSer}}
\end{figure}

\begin{figure}[!ht]
\includegraphics[angle=0,width=\textwidth,trim=30 30 30 50,clip]{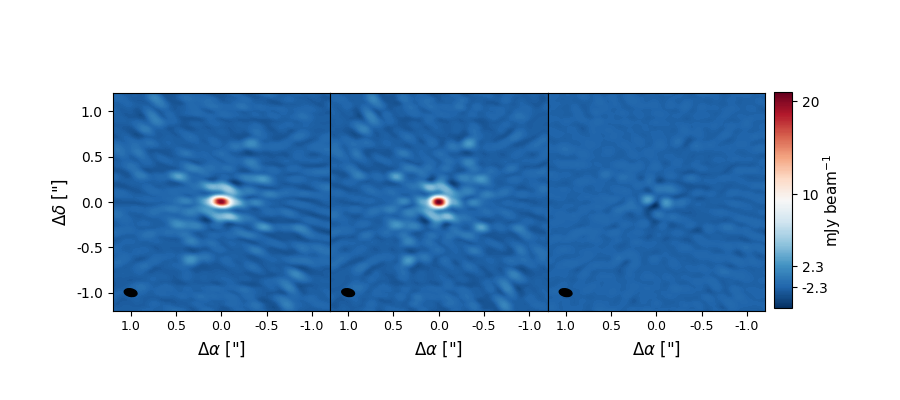}
\caption{Same as \autoref{fig:L1551} but for HBC~687.\label{fig:HBC687}}
\end{figure}

The RT modelling used an image plane fitting approach. The observed dirty maps, best fit models convolved with the dirty beam, and data minus model residuals for each system are shown in Figures \ref{fig:L1551}--\ref{fig:HBC687}.

\clearpage

\section{Posterior Distribution Functions for the RT modeling}
\label{sec:app_c}

\begin{figure}[!ht]
\includegraphics[angle=0,width=\textwidth]{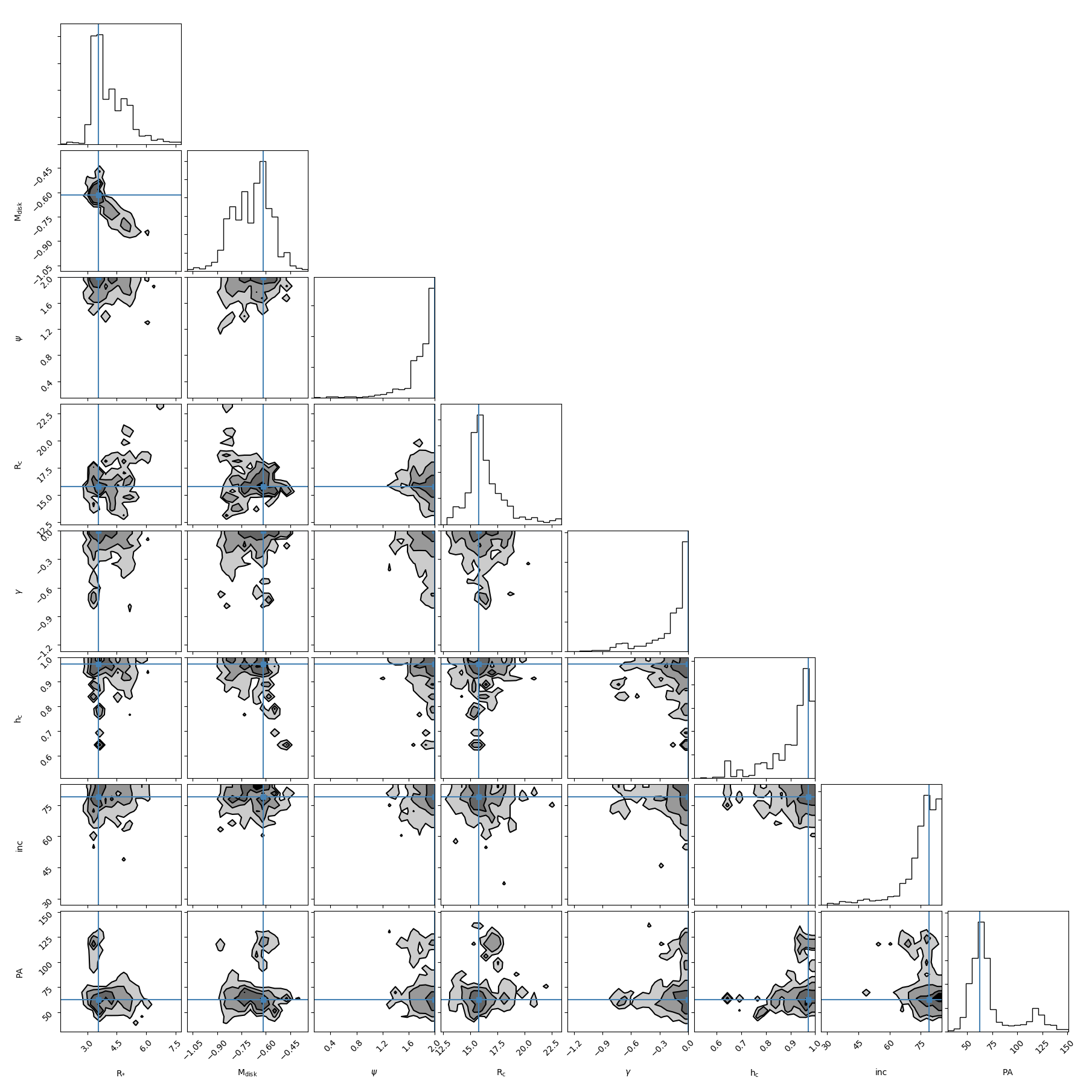}
\caption{Posterior distributions for L1551~IRS~5~N.\label{fig:L1551_N_PDF}}
\end{figure}

\begin{figure}[!ht]
\includegraphics[angle=0,width=\textwidth]{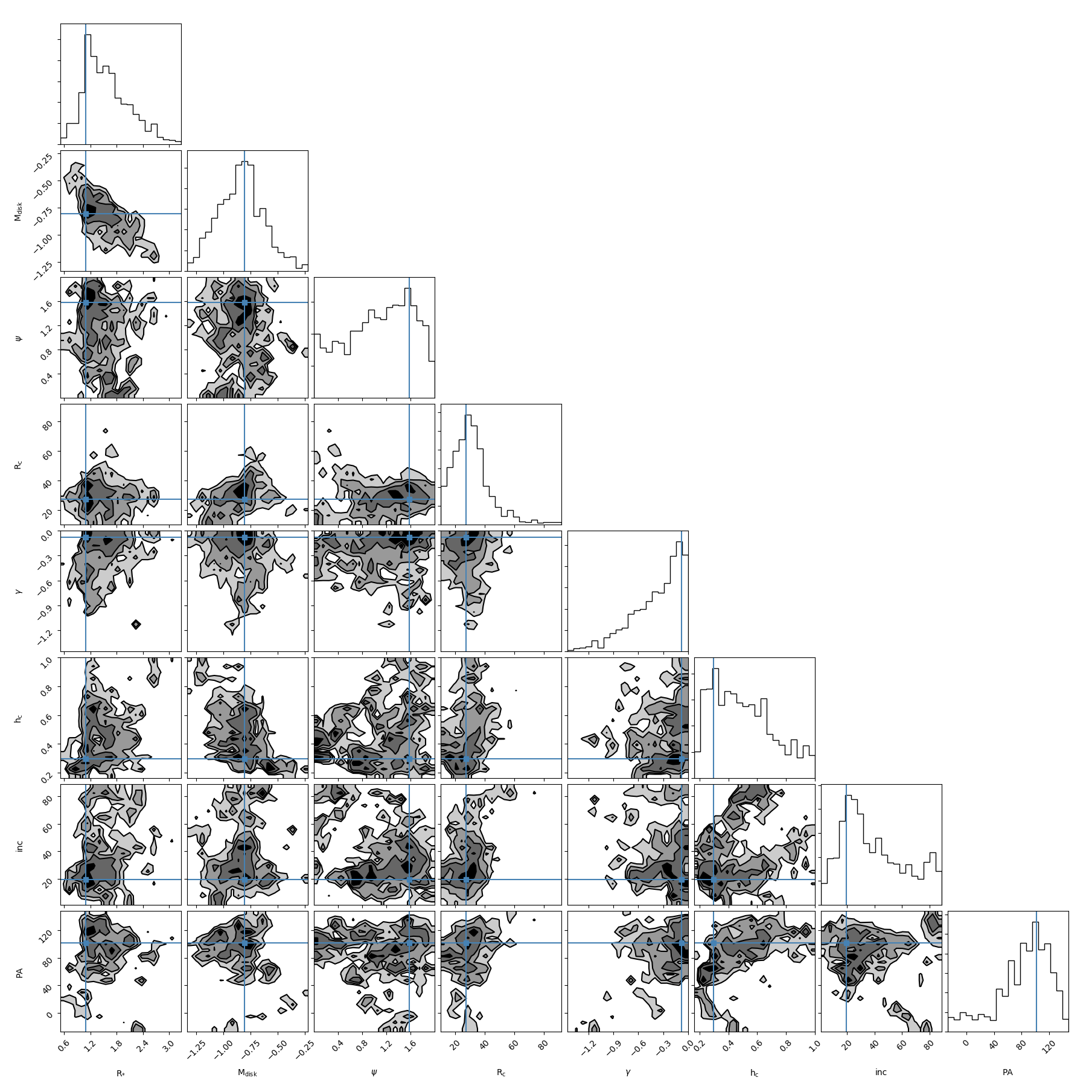}
\caption{Posterior distributions for Bran~76.\label{fig:Bran76_PDF}}
\end{figure}

\begin{figure}[!ht]
\includegraphics[angle=0,width=\textwidth]{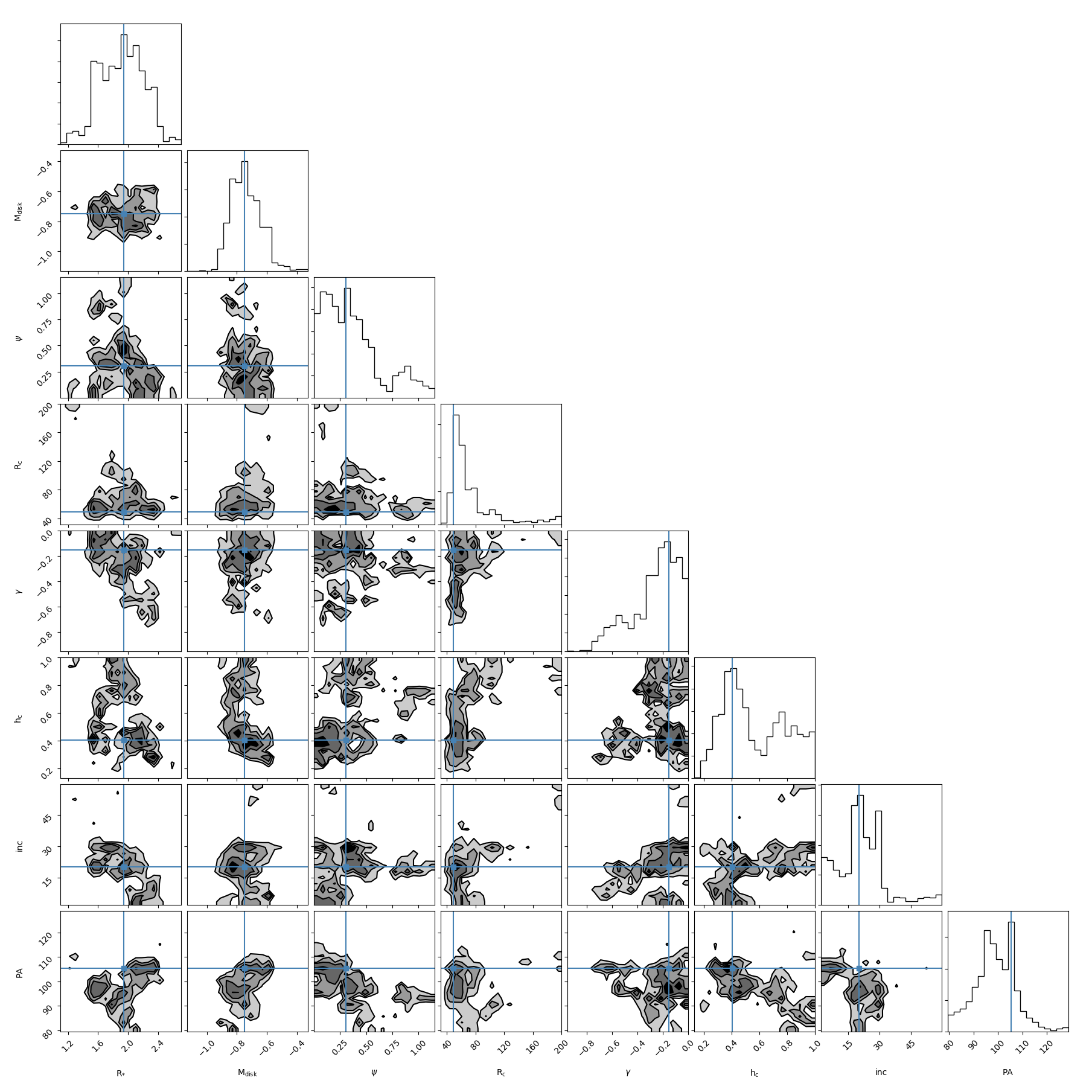}
\caption{Posterior distributions for V346~Nor.\label{fig:V346Nor_PDF}}
\end{figure}

Figures \ref{fig:L1551_N_PDF}--\ref{fig:V346Nor_PDF} show examples for the posterior distributions of the fitted parameters in the RT modeling.

\clearpage

\section{Comparison of geometrical disk parameters}
\label{sec:app_d}

\begin{figure}[!ht]
\gridline{\fig{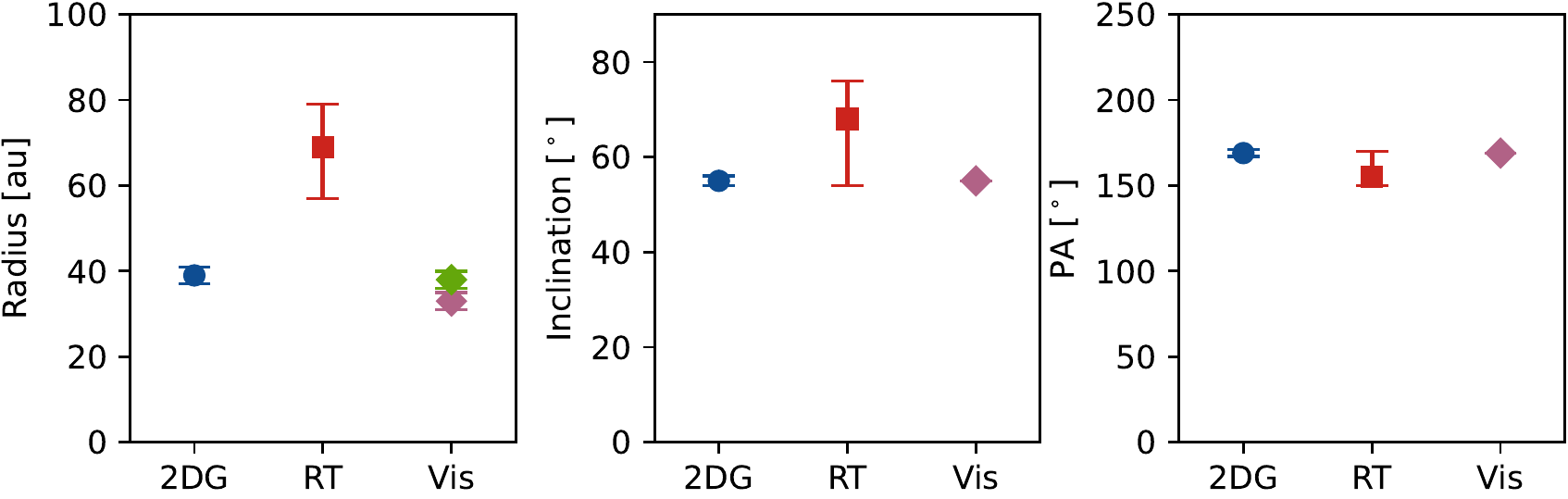}{0.45\textwidth}{(a) Haro~5a~IRS}
          \fig{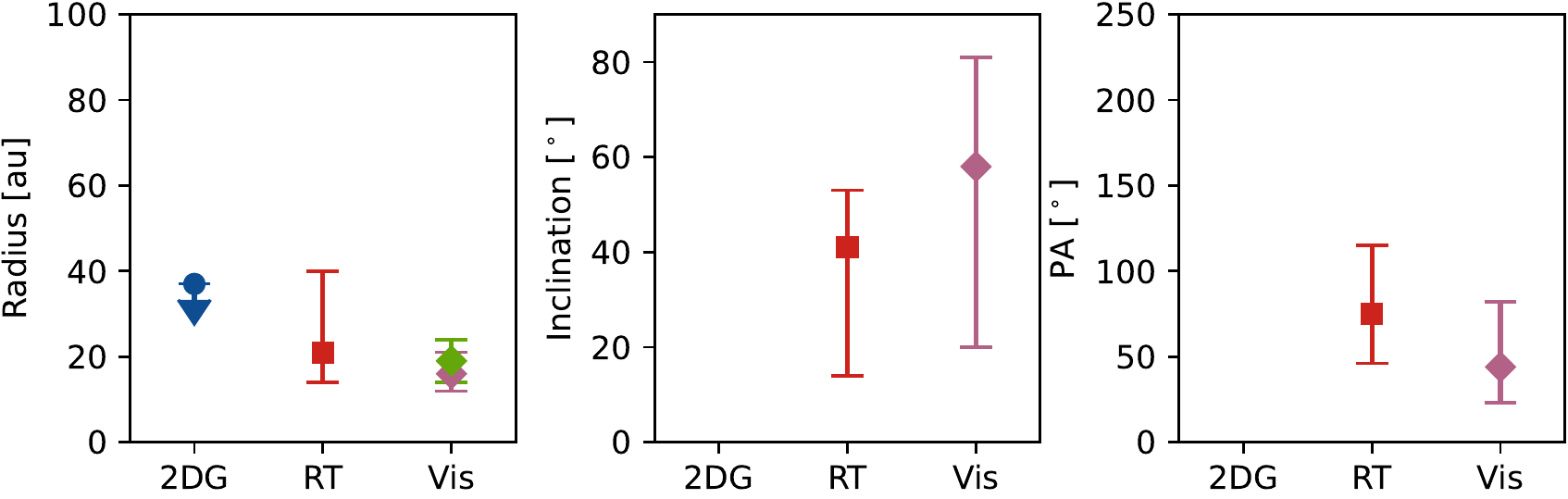}{0.45\textwidth}{(b) V899~Mon}}
\gridline{\fig{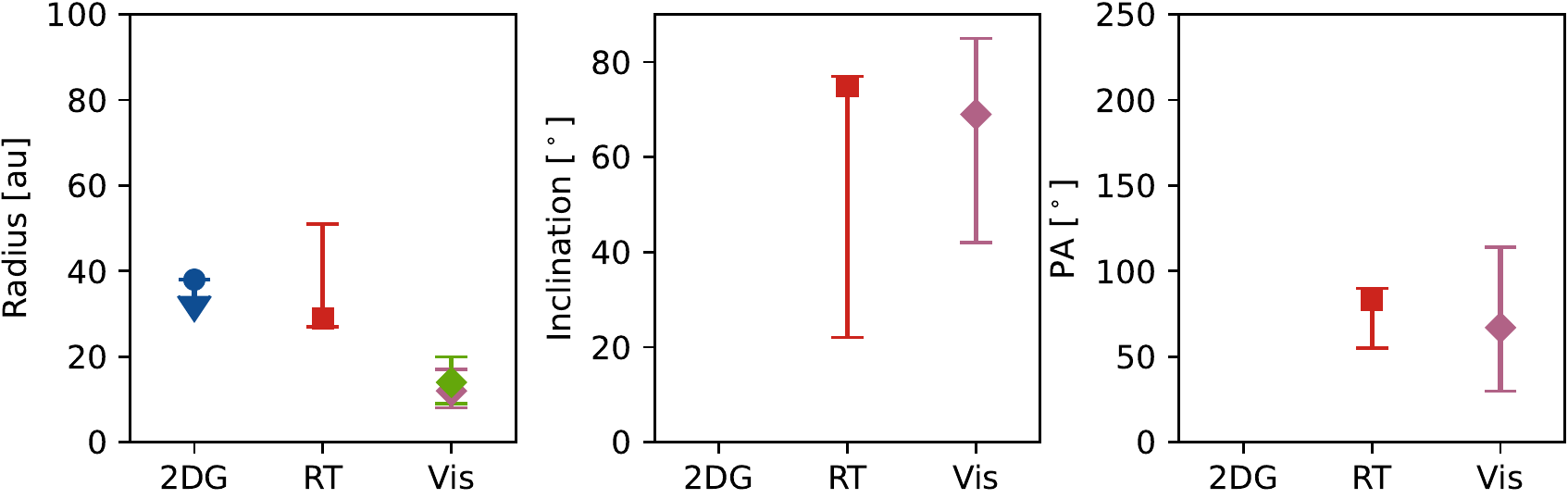}{0.45\textwidth}{(c) AR~6A}
          \fig{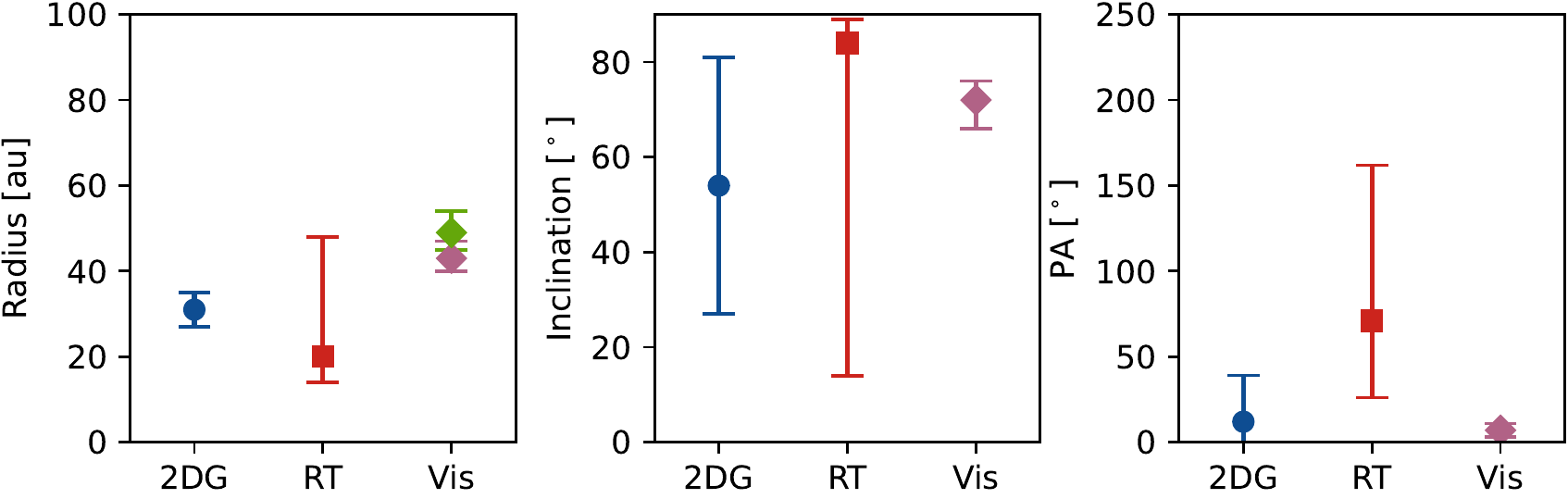}{0.45\textwidth}{(d) AR~6B}}
\gridline{\fig{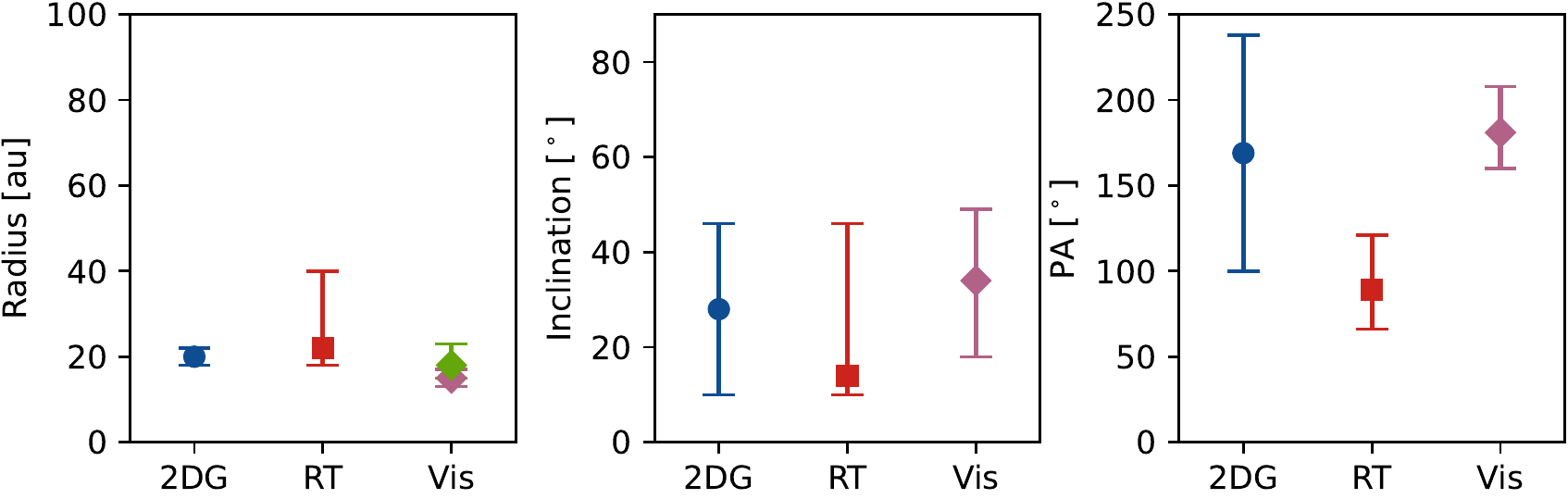}{0.45\textwidth}{(e) V900~Mon}
          \fig{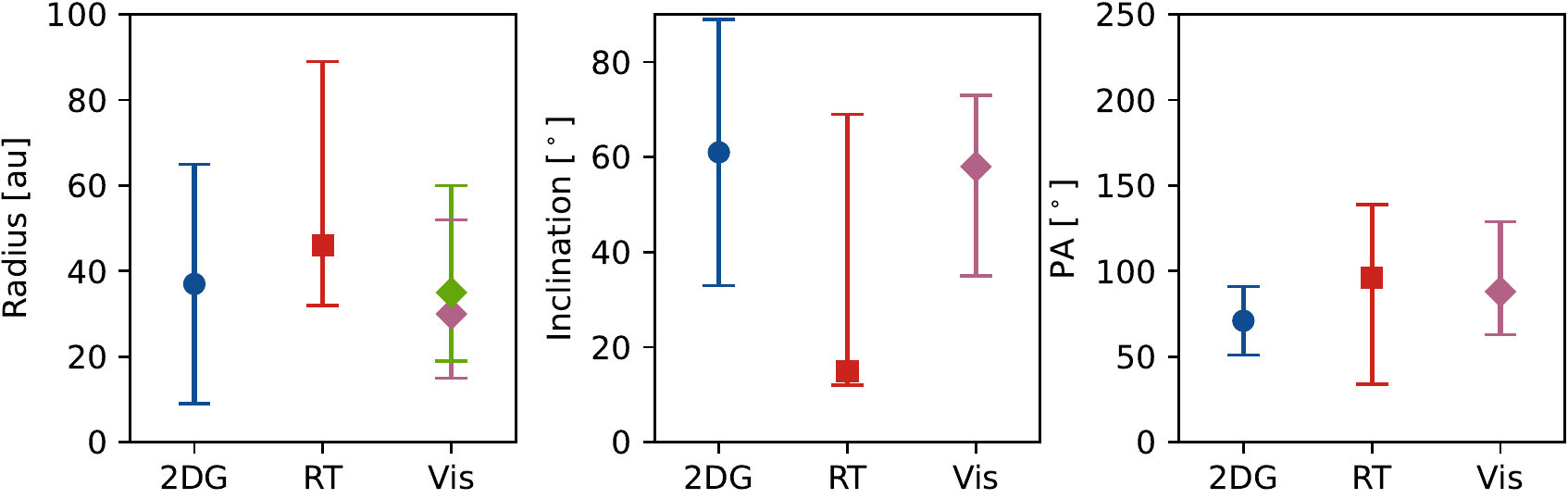}{0.45\textwidth}{(f) V960~Mon}}
\gridline{\fig{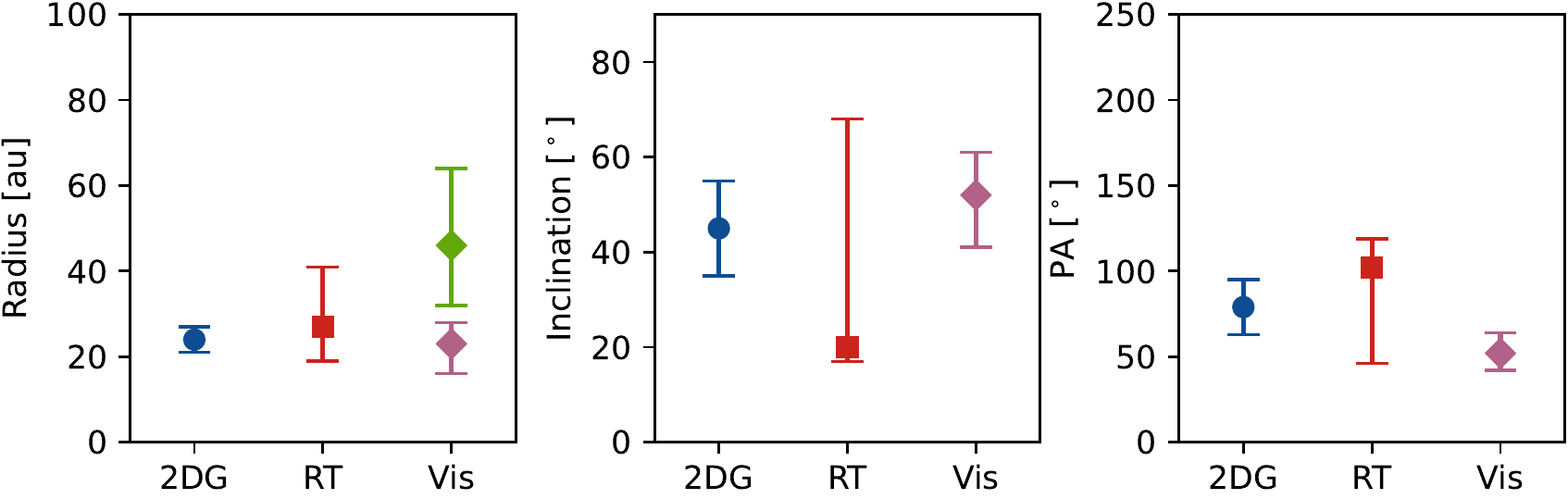}{0.45\textwidth}{(g) Bran~76}
          \fig{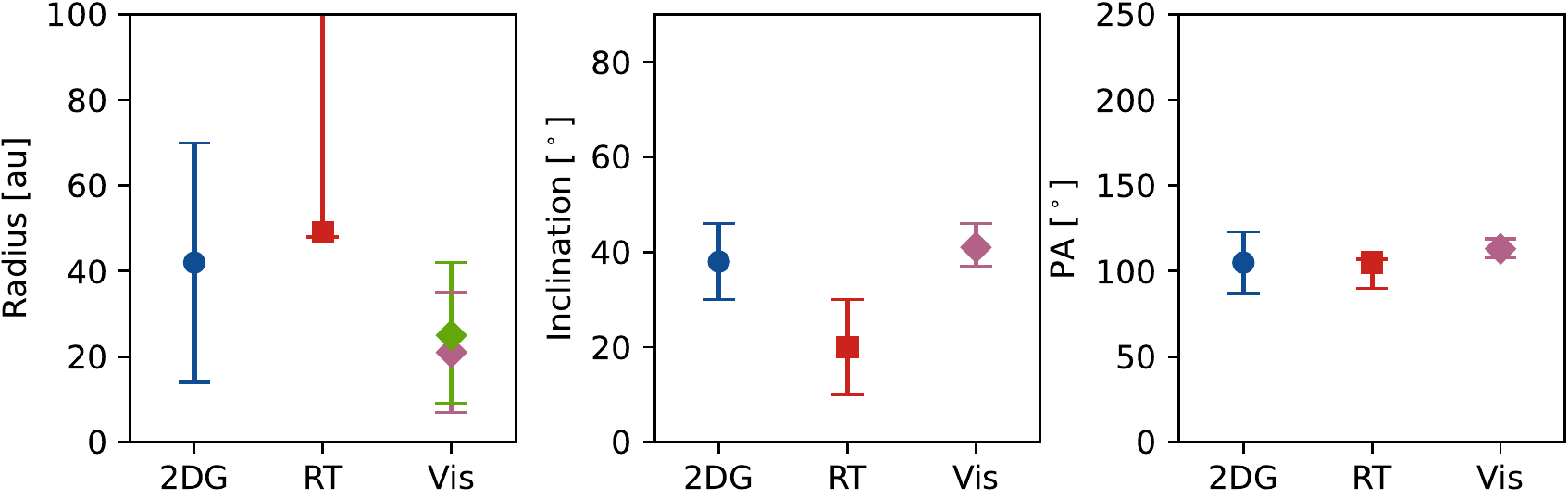}{0.45\textwidth}{(h) V346~Nor}}
\gridline{\fig{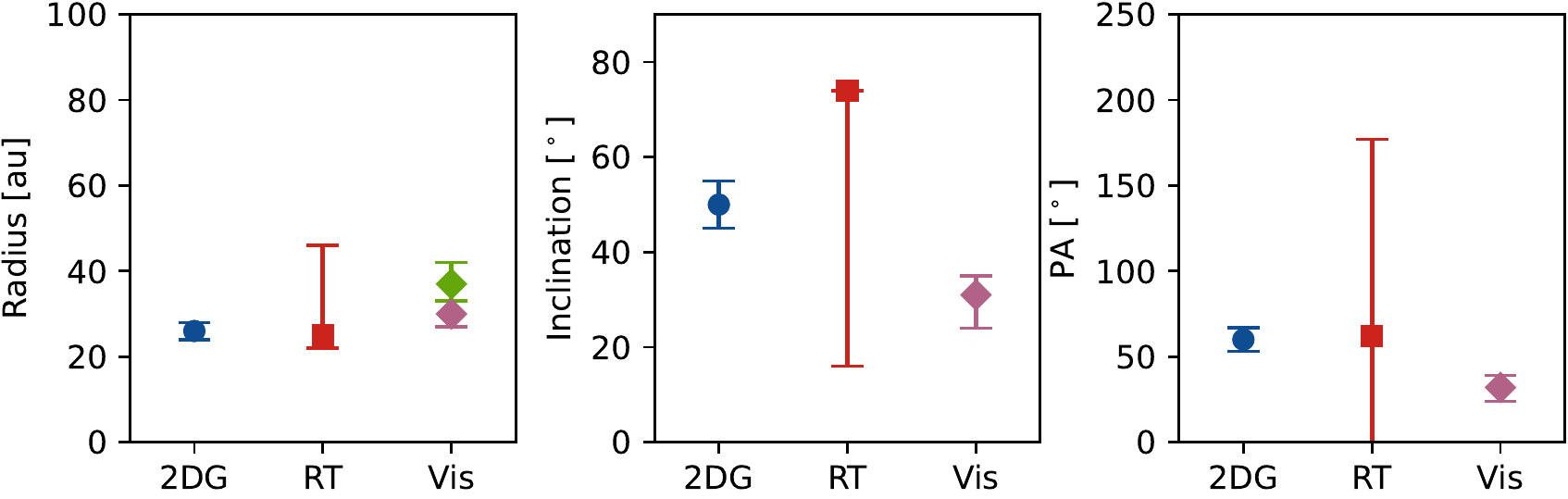}{0.45\textwidth}{(i) OO~Ser}
          \fig{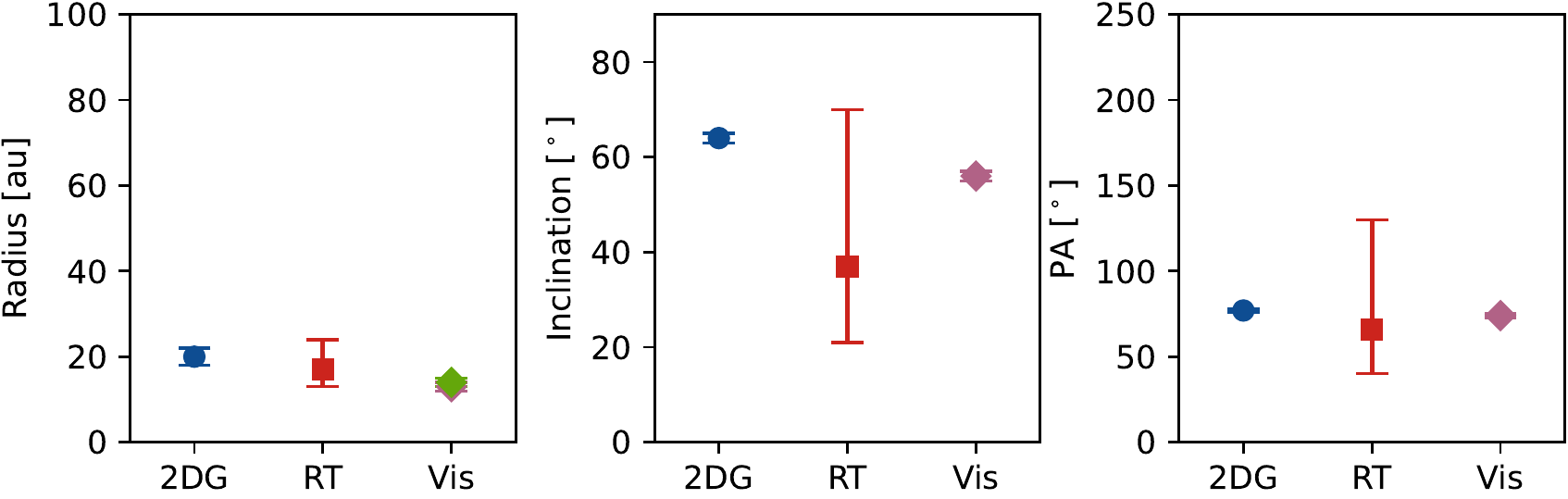}{0.45\textwidth}{(j) HBC~687}}
\caption{Comparison of the best-fitted geometrical parameters (radius, inclination and position angle) of the disks with three different methods. The blue circles show the 2D Gaussian fitting in the image space, the red squares are the results from the radiative transfer modeling in the image space with RADMC-3D, and the diamonds for disk model fitting in the visibility space. For the latter, the purple symbol indicates $R_{\rm eff}$ while the green symbol $R_{\rm disk}$. Uncertainties correspond to 1$\sigma$.\label{fig:comp_plots}}
\end{figure}

Figure \ref{fig:comp_plots} shows a comparison of the geometrical disk parameters (radius, inclination, and position angle) we obtained from the different types of modeling. 2DG stands for 2D Gaussian fitting in the image space, RT stands for radiative transfer modeling in the image space with RADMC-3D (\autoref{sec:rtmodeling}), and Vis stands for disk model fitting in the visibility space (\autoref{sec:vismodeling}).

\end{document}